\documentclass[letterpaper,twocolumn,10pt]{article}

\usepackage[margin=1in]{geometry}
\usepackage{newtxtext,newtxmath} % Times-like text & math
\usepackage[T1]{fontenc}
\usepackage[utf8]{inputenc}

\usepackage[hidelinks]{hyperref}
\usepackage{url}

\usepackage{tikz}
\usepackage{graphicx}
\usepackage{multirow}
\usepackage{booktabs} % For professional looking tables
\usepackage{pgfplots}
\pgfplotsset{compat=1.18} % ensure recent pgfplots compat
\usepackage{comment}
\usepackage{listings}
\usepackage{tabularx}
\usepackage{adjustbox}
\usepackage{array}
\usepackage{xcolor}
\usepackage{CJKutf8}
\usepackage{longtable}
\usepackage{float}
\usepackage{pifont}
\usepackage{tcolorbox}
\usepackage[toc,acronym]{glossaries}
\usepackage{soul}
\usepackage{paralist}
\usepackage{filecontents}
\usepackage{dblfloatfix}
\usepackage{balance}
\makeglossaries
\glsdisablehyper
\newacronym{api}{API}{Application Programming Interface}
\newacronym{dbi}{DBI}{Dynamic Binary Instrumentation}
\newacronym{art}{ART}{Android Runtime}
\newacronym{der}{DER}{Defined Evasion Rule}
\newacronym{jit}{JIT}{Just-In-Time}
\newacronym{drm}{DRM}{Digital Rights Management}
\newacronym{jni}{JNI}{Java native Interface}
\newacronym{ipc}{IPC}{Inter Process Communication}
\newacronym{dvm}{DVM}{Dalvik Virtual Machine}
\newacronym[plural=SDKs, firstplural=Software Development Kits (SDKs)]{sdk}{SDK}{Software Development Kit}
\newacronym{dex}{DEX}{Dalvik Executable}
\newacronym{saas}{SaaS}{Software-as-a-Service}
\newacronym{ebpf}{eBPF}{extended Berkeley Packet Filter}
\newacronym{jvm}{JVM}{Java Virtual Machine}
\newacronym{aot}{AOT}{Ahead-Of-Time}
\newacronym{ndk}{NDK}{Native Development Kit}
\newacronym{selinux}{SELinux}{Security-Enhanced Linux}
\newacronym{apk}{APK}{Android Application Package}
\newacronym{oem}{OEM}{Original Equipment Manufacturer}
\newacronym{jdwp}{JDWP}{Java Debug Wire Protocol}
\newacronym{jdb}{JDB}{Java Debugger}
\newacronym{etm}{ETM}{Embedded Trace Macrocell}
\newacronym{dds}{DDS}{Dalvik Data Structure}
\newacronym{ip}{IP}{Intellectual Property}
\newacronym{toctou}{TOC-TOU}{Time-of-Check to Time-of-Use}
\newacronym{udf}{UDF}{Unique Device Fingerprints}
\newacronym{lru}{LRU}{Least Recently Used}
\newacronym{aosp}{AOSP}{Android Open Source Project}
\newacronym{pii}{PII}{Personally Identifiable Information}
\newacronym{tdf}{TFD}{Total Device Fingerprints}

\newcommand{\subhead}[1]{\vspace {1pt}\noindent{\textbf{#1.}}}
\let\tt\texttt
\newcommand{\ourframework}{Purifire}
\newcommand{\cmark}{\ding{51}} % ✓
\newcommand{\xmark}{\ding{55}} % ✗

\lstdefinelanguage{json}{
    basicstyle=\ttfamily\footnotesize,
    numbers=left,
    numberstyle=\tiny\color{gray},
    stepnumber=1,
    numbersep=5pt,
    showstringspaces=false,
    breaklines=true,
    frame=single,
    stringstyle=\color{black},
    commentstyle=\bfseries,
    morecomment=[s]{/*}{*/},
    literate=
     *{"condition":}{{{\color{blue}"condition"}}:}{11}
      {"comm":}{{{\color{red}"comm"}}:}{7}
      {"tname":}{{{\color{red}"tname"}}:}{8}
      {"syscall":}{{{\color{red}"syscall"}}:}{10}
      {"args":}{{{\color{red}"args"}}:}{7}
      {"data":}{{{\color{red}"data"}}:}{7}
      {"where":}{{{\color{red}"where"}}:}{8}
      {"evasion":}{{{\color{blue}"evasion"}}:}{10}
}
\title{To Unpack or Not to Unpack:\\
Living with Packers to Enable Dynamic Analysis of Android Apps}

\author{
  Mohammad Hossein Asghari \\
  Carleton University \\
  \texttt{hosseinasghari@cmail.carleton.ca}
  \and
  Lianying Zhao \\
  Carleton University \\
  \texttt{Lianying.Zhao@carleton.ca}
}

\begin{document}
\date{}
\maketitle

% =========================

\begin{abstract}

%-------------------------------------------------------------------------------
\small
Android apps have become a valuable target for app modifiers and imitators due to its popularity and being trusted with highly sensitive data. Packers, on the other hand, protect apps from tampering with various anti-analysis techniques embedded in the app. Meanwhile, packers also conceal certain behavior potentially against the interest of the users, aside from being abused by malware for stealth. Security practitioners typically try to capture undesired behavior at runtime with hooking (e.g., Frida) or debugging techniques, which are heavily affected by packers.
%were left with no choice other than unpacking the packed apps which are short-lived and require Android instrumentation that requires Android technical knowledge. Additionally, 
Unpackers have been the community's continuous effort to address this, but due to the emerging \emph{commercial} packers, our study shows that none of the unpackers remain effective, and they are unfit for this purpose as unpacked apps can no longer run. %On top of that, packers detect anomalies like dynamic analysis tools fingerprints (debuggers like JDB or DBIs like frida) and often terminate or crash. 

We first perform a large-scale prevalence analysis of Android packers with a real-world dataset of 12,341 apps, the first of its kind, to find out what percentage of Android apps are actually packed and to what extent dynamic analysis is hindered. We then propose \textit{{\ourframework}}, an evasion engine to bypass packers' anti-analysis techniques and enable dynamic analysis on packed apps \emph{without unpacking them}. {\ourframework} is based on eBPF, a low-level kernel feature, which provides observability and invisibility to user space apps to enforce defined evasion rules while staying low-profile. %{\ourframework} enables hooking and debugging of packed apps for security practitioners to use powerful and popular tools like Frida.
Our evaluation shows that \ourframework{} is able to bypass packers' anti-analysis checks and more importantly, for previous research works suffering from packers, we observe a significant improvement (e.g., a much higher number of detected items such as device fingerprints).
\end{abstract}

\section{Introduction}

Mobile applications (apps) are now a key part of daily life, supporting communication, shopping, entertainment, and many other activities. Android has the largest market share worldwide (71\%) \cite{turner2025android}, which also makes it a target for attackers who try to reverse-engineer or modify apps to steal intellectual property, remove restrictions, or commit fraud. To protect against these threats, Mobile Application Security Verification Standard (MASVS) \cite{owasp_masvs} dictates security guidelines for Android developers to build their app based on standard self-protection techniques to protect the app from being tampered with, referred to as anti-analysis techniques hereafter.
Over the decades, software communities, both in open‑source and commercial domains, have relied on methods like code obfuscation and encryption in desktop environment such as Linux and Windows to hinder reverse engineering and analysis \cite{ugarte2015sok}. These methods inspired the more formal notion of “packing”, where programs are transformed  to conceal their content or obfuscated to hinder analysis, often used both for legitimate hardening and malware evasion \cite{cheng2021obfuscation, bayer2009view}. Android packers adopt concepts from desktop packers and, by leveraging the unique structure and execution model of Android apps, provide similarly sophisticated packing features. These packers use encryption, code obfuscation, and other anti-analysis techniques to protect app code. While they help developers protect their work, these protections also make security analysis harder, and they can be misused by malware authors to hide harmful behavior \cite{Duan2018ThingsYM, wong2018tackling}.

Packers decrypt and load app code at runtime, add dependencies, and use multiple protection techniques such as anti-debugging, anti-instrumentation, and environment checks. In addition to IP protection, packed apps actively respond to anomalies that trigger their runtime anti-analysis mechanisms. These mechanisms may scan the file system for specific artifacts or inspect process memory for suspicious values. When such conditions are detected, the app may terminate gracefully or intentionally crash to prevent further analysis \cite{ibrahim2021safetynot,haupert2018honey,kondracki2022droid}. Additionally, packers can detect and block popular dynamic analysis tools like Frida \cite{frida_website} and Xposed \cite{lsposed_github}, which are often used for runtime inspection, finding vulnerabilities, and checking for privacy risks \cite{ruggia2024unmasking}.

Previous approaches to analyzing packed apps, such as unpacking, proved effective at the time \cite{dexx2018sun, zhang2015dexhunter, blackdex_github, appspear2015yang}, but their efficacy has diminished over time as packers have evolved. Even if unpacking works, the recovered Dex code is often incomplete~\cite{Duan2018ThingsYM, packergrind2017leixue}, obfuscated, dependent on packer code due to on-demand dynamic code loading and multi-layer packing~\cite{ugarte2015sok}, or at least missing metadata. As a result, obtaining only dumped Dex code still leaves it very challenging, if not impossible, to repackage and analyze the app dynamically at runtime. Furthermore, numerous modern apps now execute mostly in native code for performance reasons~\cite{wong2018tackling} limiting the usefulness of dumped Dex code.

Nowadays, commercial packers provide developers with accessible, zero-code solutions for app hardening with different subscription levels (see Section~\ref{packing-features}). Well-known commercial packers include Chinese services such as Ijiami \cite{iJiami}, Qihoo \cite{360Dev}, Tencent \cite{TencentCloud}, Bangcle \cite{Bangcle}, and Baidu \cite{BaiduApp}, and non-Chinese solutions such as Appdome \cite{appdome}, LIAPP \cite{liapp_website}, and Dovern \cite{doverunner}. These services accept APK files as input and return a packed version of the app. The advent of commercial packers almost rendered the goal of defeating %packed apps 
packers
unrealistic.

%Considering infeasibility of fully unpacking the packed apps, the ongoing struggle between protection methods and analysis tools means researchers need more practical solutions.

We propose a different approach. Instead of trying to fully defeat packers, we aim to live alongside with them by enabling dynamic analysis without unpacking the app. Our framework, called \ourframework, uses the Extended Berkeley Packet Filter (eBPF)~\cite{ebpf_website}, a kernel feature to run privileged code in a safe and event-driven manner,  to intercept and bypass anti-analysis checks while the app is running without being detected. This allows tools like Frida to work on packed apps based on what is available at runtime without being detected. Purifire employs a configurable evasion system that enables analysts to bypass anti-analysis techniques and examine both benign and malicious apps. %while preserving the packer’s IP protections. %Our approach can enable other Frida-based academic tools that were previously blocked by packers (see Section~\ref{evaluations}). 
In addition, we conduct a large-scale study of commercial packer usage in Android apps to measure their prevalence and perform dynamic analysis to identify implemented anti-analysis techniques. 
This is to find out the actual impact of Android packers and to what extent security analysis can be affected. 
We evaluate \ourframework{} by contrasting how several academic security projects previously blocked/affected by packers with the improvements after \ourframework{} is used (see Section~\ref{evaluations}). 

\subhead{Contributions} %The main contributions of this paper are as follows.
\vspace{-5pt}
\begin{itemize}
\itemsep0em 
    \item %\subhead{Prevalence Study of Packers} 
    We conduct a large-scale prevalence analysis of Android packers over 12,341 real-world apps -- 7,913 from Chinese app stores and 4,428 global, to measure the actual impact of commercial packers. Our findings reveal that nearly 50\% of Chinese apps use packing services (while non-Chinese apps see a much lower percentage), and that existing analysis tools are largely ineffective against these protections. To the best of our knowledge, this is the first systematic prevalence study on this scale.
    \item We examine the technical aspects of Android packers and unpacking techniques with a taxonomy, highlighting their impact on dynamic program analysis. This serves as a foundation for future work in understanding and mitigating packing-related challenges.
    \item We propose and implement {\ourframework}, the first kernel-level eBPF-based evasion framework designed to bypass anti-analysis defenses without unpacking the app. Purifire enables dynamic analysis tools to operate transparently even on protected apps, offering a practical and customizable solution for security analysts.
\end{itemize}
%-------------------------------------------------------------------------------

\section{Motivation}

%\st{Even when state-of-the-art unpackers succeed, the recovered Dex is frequently incomplete, entangled with packer runtime logic, and irrelevant as modern apps shift critical logic to native code and multi-stage, on-demand loading. Not to mention the extracted Dex has already lost its metadata. In practice, the result is code you cannot repackage and run (sometimes called de-shelling), so the runtime behavior cannot be observed.}
Our findings indicate that a significant share of apps are affected by packers (see Section~\ref{prevalence-analysis}), showing the importance of being able to analyze packed apps as they are becoming common. Many existing tools and academic studies are built on ready-to-use well-maintained tools such as Frida and Xposed scripts (see Section~\ref{frida-usecase}). However, their effectiveness is often reduced by packers, which detect them and react.

Unpacking once promised a path to visibility, but it no longer delivers what dynamic analysis actually needs. To confirm this, we applied available unpacking approaches to our dataset to evaluate their effectiveness (see Section~\ref{unpackers-tests}) and found that they failed to recover the majority of the apps, primarily due to their fixed configurations and outdated design. This is further worsened by the growth of commercial Android packers with sophisticated techniques never fully exposing clear code at runtime (but on-demand), invalidating the design of most unpackers.

Meanwhile, commercial packers actively evolve and constantly add new unpacker “fingerprints”, which reflects the nature of an arms race. Our manual analysis shows packers like Ijiami \cite{iJiami} check for unpacker configuration files, path, and package names like: \texttt{/data/fart} (FART~\cite{fart_github}), \texttt{/data/local/tmp/unpacker.config} (Youpk~\cite{unpacker_github}), \texttt{/data/dexname} (DexHunter~\cite{zhang2015dexhunter}), and \texttt{top.niunaijun.blackdex} (BlackDex~\cite{blackdex_github}).

%When {\ourframework} is integrated into their workflow, we observe a marked improvement in results, demonstrating that our evasion engine can significantly enhance their analysis capabilities (see Section~\ref{evaluations}).

\iffalse
\textcolor{purple}{START: I moved the Unpacker lack of evaluation here without mentioning malware Remove it if points already covered!}

Most unpacker evaluations suffer from limited or unclear ground truth. Several studies report results only on small sets of self-packed benign or F-Droid apps where the original Dex is known (e.g., DexHunter~\cite{zhang2015dexhunter}, AppSpear~\cite{appspear2015yang}, PackerGrind~\cite{packergrind2017leixue}). Others target wild samples but cannot measure the total number of classes or methods, since the true Dex code is unknown, dynamically released at runtime, or have corrupted headers (e.g., Gupacker~\cite{zheng2025gupacker}, BPFDex~\cite{li2025bpfdex}). Some works instead report success by producing a parseable Dex file or higher method coverage without verifying full code recovery (e.g., DexX~\cite{dexx2018sun}, ReDex~\cite{cai2020redex}). In addition, most papers do not clarify whether they used basic or advanced commercial packer services (see Section~\ref{commercial_packer}), leaving uncertainty about dataset representativeness. These issues highlight a methodological gap: unpacker evaluations often rely on indirect indicators rather than quantifying how much of the original app code is truly recovered.

\textcolor{purple}{END}
\fi

On the other hand, unpackers often rely on \gls{art} instrumentation and emulation-based approaches, both of which require advanced technical expertise, and are
%. These methods are impractical for large-scale dynamic analysis by security analysts because most are 
quickly outdated and based on modified \gls{aosp}, which is incompatible with Play Integrity \cite{play_integrity_api} and Google Services. In practice, analysts often use a real phone with a factory image to remain low-profile, minimize the risk of triggering anti-analysis techniques, and observe the app’s complete functionality, including interactions with Google Services (e.g., Firebase \cite{firebase}).
%\textcolor{purple}{CHECK HERE added from unpacker obstacles}
%In contrast, unlike unpacking, hooking provides a more efficient and effective alternative. 
Their common practice involves the aforementioned tools like Frida, which support app-specific instrumentation scripts to trace class and method invocations, such as monitoring the transformation of network packets. These tools enable hybrid analysis approaches that combine runtime visibility with targeted inspection, making them indispensable for modern Android app analysis.

The infeasibility of unpacking and the need for preserving well-established dynamic analysis tools call for a paradigm shift toward analyzing packed apps without unpacking them, as with malware analysts relying on debuggers to examine behavior in-place.
%Therefore, the security community needs to find a balance. App analysts should be able to analyze packed apps without fully unpacking the app. 
To support this goal, we propose an evasion engine that enables controlled, single-stepping analysis via predefined evasion rules. This approach helps security analysts study packed apps at runtime, while still respecting the protections applied by developers to avoid outright code theft in its entirety.

\iffalse
In this paper, we aim to find out the answer to the following research questions:
\begin{itemize}
    \item \subhead{RQ1} How do commercial Android packers defeat current dynamic analysis methods?
    \item \subhead{RQ2} What is the current prevalence of packers among Android apps?
    \item \subhead{RQ3} What are the technical choices for both packing efforts and unpacking efforts?
    \item \subhead{RQ4} If commercial Android packers are hard to defeat (and a total defeat is not necessarily a goal), can we still enable limited dynamic analysis by evading anti-analysis techniques at runtime?
\end{itemize}
\fi

\subsection{Threat Model}
In the Android app ecosystem, multiple stakeholders interact with different levels of knowledge and control. App developers create features and may rely on packing services to protect their code/data, while app users only care about the final functionality. Security analysts aim to evaluate app security without access to source code or packer internals, and in our discussion context, evasion rule authors develop rules to bypass anti-analysis methods %without full knowledge of every packed app’s behavior.
(e.g., via reverse-engineering).
Packing services, %although intended to protect apps, can also be exploited by 
mostly driven by economic incentives, 
intend to achieve the protection goals by constantly improving packers.
Against the interest of app users (and potentially other entities), while 
malicious actors can abuse packers outright to conceal harmful code, 
certain behavior introduced by app developers in the first place, intentionally or inadvertently, may also remain unnoticed due to packers.

%making it difficult to distinguish between legitimate and malicious use. This creates an environment where protective tools can also serve as tools for evasion.

These dynamics create a complex landscape. Mobile apps often receive more trust than PC software due to Android’s sandboxing and permission control, allowing sensitive operations that would be risky elsewhere. However, the same isolation that shields users can also hide undesired behavior, with no single actor with complete control or visibility.
%the threat becomes a shared and evolving challenge. %To address this, the security community must strike a balance: enabling minimal security analysis of packed apps. 
Our proposed evasion engine 
aims to achieve a balance between stakeholders by
avoiding fully unpacking apps (hence defeating packers) and
allowing security analysts to 
observe app behavior at runtime.
%supports controlled, step-by-step analysis, preserving app protections while allowing effective security evaluation.

\section{Background}

To facilitate subsequent discussions, we first briefly discuss how Android app security analysis benefits from \gls{dbi} tools, and the uniqueness of commercial Android packers.

%to examine an effective dynamic analysis method via tools like Frida. We also review studies that employ Frida and Xposed as the core components of their analysis frameworks and highlight the significance of packers’ ability to detect such tools.% Furthermore, we compare Android with desktop systems, discussing how Android-specific packers can introduce a new phase in packer analysis and emphasizing the importance of examining their unique characteristics (see~\ref{android-uniqueness}).%
%Finally, we examine packing services, assessing their accessibility to developers and summarizing the features they offer based on information available on their websites (see Section~\ref{commercial_packer}).

\subsection{Android Dynamic Binary Instrumentation} \label{frida-usecase}
Dynamic analysis is widely used to study Android apps, as it helps bypass runtime decryption and obfuscation that hinder static analysis~\cite{barros2015static}. Frameworks such as Frida~\cite{frida_website} and Xposed~\cite{lsposed_github} allow analysts to hook \gls{api} calls, modify parameters or return values, replace method implementations, intercept network traffic, audit cryptographic functions, and trace control flow. However, despite their popularity in academic research, these \gls{dbi} tools are highly detectable by modern packers, which implement checks for tracing, runtime modifications, and instrumentation artifacts. As a result, they are often ineffective against packed apps, and the rapid evolution of Android and packer techniques creates a heavy maintenance burden. In the following, we provide a few examples of the use of \gls{dbi} tools for better security/privacy, and the foreseeable benefit of re-enabling them in the face of packers.

%Possemato et al.~\cite{possemato2020towards} hooked the network \gls{api} \tt{Socket.connect} using Frida. Wang et al.~\cite{wang2021understanding} and the ZjDroid project~\cite{halfkiss_zjdroid} employed Xposed to capture network traffic, although Xue et al.~\cite{xue2017malton} noted that Xposed can be detected via \textit{Intent Registration}.

Frida can be used for file and memory operations. Dong et al.~\cite{dong2024exploring} hooked file-related APIs such as \texttt{java.io.File} and \texttt{open} from \texttt{libc} to conduct a large-scale study of tracking SDKs using external storage as a covert channel, and Pourali et al.~\cite{certvalidation2024Sajjad} retrieved stack traces to capture insecure TLS certificate validation. Anglano et al.~\cite{anglano2023enabling} hooked popular encryption libraries (SQLCipher, Realm and Jetpack Security) for forensic analysis. %Mixon et al.~\cite{mixon2025hidden} traced critical libraries and functions identified during static analysis, and used fridump~\cite{fridump_nightbringer21} to extract sensitive data such as credentials from process memory. 

Frida has also been used for \gls{api} monitoring. Heid et al.~\cite{heid2024haven} modified device properties to study their effect on generated device fingerprints. Ibrahim et al.~\cite{ibrahim2021safetynot} examined the SafetyNet attestation \gls{api} through Frida hooks. %In~\cite{droidkex}, Frida was used to intercept network traffic and recover the \textit{MasterSecret} of each connection. 
Cui et al.~\cite{Cui2022TraceDroid} identified privacy leakage from third-party libraries, while Aldoseri et al.~\cite{aldoseri2022insecure} intercepted URI handling in the top 15 popular mobile browsers and discovered four CVEs. 

Additionally, Diamantaris et al.~\cite{diamantaris2019reaper} utilized Xposed for real-time analysis of permission checks and stack trace monitoring in Android apps. %Continella et. al~\cite{continella2017obfuscation} implemented a module, based on Xposed, which hooks method calls and records and modifies their return values and bypass SSL pinning via JustTrustMe module~\cite{justtrustme_fuzion24}.
Moreover, community-developed frameworks with embedded Frida-based scripts are widely used in app analysis~\cite{zerbini2024r, koch2023ok}, including tools such as Medusa~\cite{medusa_ch0pin} and Objection~\cite{objection_sensepost}.

% \textcolor{red}{not effecting packed apps but interesting use-case for frida}

% There are also several studies that use frida to bypass runtime anti-analysis techniques like root detection \textcolor{red}{Android reference needed for root detection}.

Several tools have been developed to support dynamic analysis, including JNITrace~\cite{chame1eon_jnitrace} for intercepting \gls{jni} interactions, Blutter~\cite{blutter_github} for analyzing apps that have been developed by Flutter~\cite{flutter_website}, and Google’s Frida-based SSL logger~\cite{geffner_ssl_logger}. These examples highlight the widespread adoption of \gls{dbi} tools, within the Android security community and the potential consequence of commercial packers rendering them ineffective.

%considering that packers detects these tools means many capabilities of dynamic analysis will be unavailable.

\subsection{Commercial Android Packers} \label{commercial_packer}

Commercial packers can be considered to be packers-as-a-service, aiming to enable developers to secure their code without requiring expertise~\cite{Yang2024BeyondHorizon} in code obfuscation or anti-analysis techniques, unlike most open-source ones. These online packing platforms offer Android app hardening, by just taking uploaded APK files as input. %making security more accessible to developers. %As a result, developers with limited technical knowledge of Android app hardening techniques, such as code obfuscators and anti-analysis security packages, can protect their apps by simply uploading them to a third-party packing service, which we refer to as a commercial packer.

%There are several ways to categorize different packers based on their protection techniques. We will mention some of their advanced techniques as listed on their websites. However, it is also important to consider the origin of these packers to better understand their usage and the products they are applied to.

%Additionally, when considering different corporations and vendors, the same app version may appear in various stores with minor modifications, often packed using different packers for each version \cite{wang2018beyond}. 

%In this section, we start by examining the registration requirement to better understand the ecosystem and then, we explore the various techniques such commercial packers advertise on their websites.

\subsubsection{Packing features} \label{packing-features}

%Packing services offer a variety of security hardening technologies and techniques to help developers protect their apps. They 

Packing services usually provide multiple tiers of protection, including free and professional (paid) plans. As a result, the same app version can exist in multiple packed variants, depending on the chosen protection level and distribution platform (e.g., various app stores other than Google Play)~\cite{beyondgoogle2018wang}.
%Commercial packing services provide layered security features to protect Android applications. 
Chinese providers such as Tencent~\cite{TencentCloud} and Manxi~\cite{Manxi} commonly market their products under terms like “reinforcement” or “shielding” while non-Chinese services like LIAPP~\cite{liapp_website} use more direct terminology. Understanding %the protections they offer not only reveals how developers can secure apps but also helps researchers identify ways to bypass these defenses.
how packing features vary with tiers/plans can make our evasion design more informed and targeted.

\subhead{Basic/free plans} Basic plans generally include well-known anti-analysis measures. Tencent and Manxi offer protections such as Dex code encryption, tamper checks, anti-dumping, and anti-debugging, while LIAPP provides more detailed options even at the basic tier. Notably, LIAPP’s plan includes game engine protection for Unity~\cite{unity_mobile}, Unreal~\cite{unreal_mobile}, and Cocos~\cite{cocos_website}, reflecting broader support for %applications in gaming environments.
game apps.

\subhead{Advanced plans} Advanced plans introduce more sophisticated defenses, often tied to paid subscriptions. Chinese packers implement techniques like DexVirtualization, DexToCPP conversion, and \texttt{AndroidManifest.xml} tamper-proofing. Tencent extends protection to native libraries with obfuscation and encryption, while Manxi offers custom native loader (see Appendix~\ref{app:native-loader}). Beyond code hardening, advanced features also include root and emulator detection, VM checks, APK/resource tamper-proofing, SSL pinning, hook detection, VPN/overlay detection, JDWP checks, and anti-bot measures. These advanced protections highlight the increasing complexity of Android-specific anti-analysis strategies.

\subsubsection{Registration requirements}

%Packing services have been widely used by app developers because they offer ready-made protection solutions, eliminating the need to create custom app-hardening schemes \cite{Yang2024BeyondHorizon}.
%However, these services were originally designed. 
Chinese packing services are region-specific, requiring mainland phone numbers, real-name authentication~\cite{liu2024tickets}, and sometimes government ID verification, CAPTCHAs, or VPN access (if accessed outside mainland China), creating a tightly controlled ecosystem.  Developers are often asked to specify the category of their app, which suggests that different protection techniques may be applied based on the app type (e.g., finance, sports)~\cite{dong2024exploring}. In contrast, non-Chinese services are more accessible, typically requiring only an email for registration. These differences between Chinese and non-Chinese packing services highlight the varying levels of restriction and protection (e.g., as a black-box in a closed ecosystem).

\iffalse
Our examination of packer websites revealed that Chinese packing services are
originally designed to cater to specific regions and 
require users to register with a mainland China phone number, making it necessary to have a physical SIM card (also involving real-identity verification as per current regulations), as virtual numbers are not accepted. 
Furthermore, certain Chinese packing services impose an extra layer of authentication called ``Real Name Authentication''~\cite{liu2024tickets} which is usually completed through a government app and linked to the user's national ID card.
In contrast, non-Chinese packing services are more accessible, typically allowing registration with an email address.

% \input{Chapter4/Figs_Tables/packer_services_registration}

The registration process for certain Chinese packing services includes additional authentication steps, such as entering verification codes received via SMS and solving CAPTCHA challenges. These CAPTCHAs often require Chinese language translation, and some services may even need VPN access due to geographical restrictions. After signing in, users are provided with a portal to upload their apps. Packing services often ask developers to specify the category of their app, which suggests that different protection techniques may be applied based on the app type (e.g., finance, sports)~\cite{dong2024exploring}.

These differences between Chinese and non-Chinese packing services highlight the varying levels of restriction and protection (e.g., as a black-box in a closed ecosystem).

\fi

\subsection{Extended Berkeley Packet Filter (eBPF)}
\Gls{ebpf}~\cite{ebpf_website} is a powerful Linux kernel feature that, beyond its origins in packet filtering, now supports system introspection, performance monitoring, and security enforcement. With features like maps, BTF type metadata, and probes (kprobes and uprobes)~\cite{rice2023learning}, it enables efficient data sharing, cross-version compatibility, and dynamic instrumentation in both kernel and user space. The CO-RE (Compile Once, Run Everywhere) capability further allows portability without recompilation, and using Aya-Rust~\cite{aya_rs_website}, our work is shipped as a statically linked \gls{ebpf} program for arm64 Android.

%-------------------------------------------------------------------------------

%\section{Existing Work and Experimental Findings}
\section{Packing and Unpacking Techniques}
%Packers protect the app's code by encryption but they also have to protect themselves from analysis tools to not exposing the encryption routine of app's data. Packers use anti-analysis techniques and obfuscation to hide their tricks and make the reverse engineers' life harder. 
In this section, we systematically analyze and survey
%studied tricks and existing techniques that packers utilize and we also provide an overview of anti-analysis principle to show how the packer can retrieve environment-related and context-related information to detect anomalies. 
currently known anti-analysis techniques and state-of-the-art unpacking approaches.
\subsection{Anti-analysis Techniques} \label{anti-analysis-tricks}
%Anti-analysis techniques have been widely studied in the literature. Many of these 
Most anti-analysis techniques originated from PC packers but have been adapted to the Android ecosystem by leveraging Android’s unique architecture. In the following, we discuss several to facilitate subsequent discussions.

%In this section, we focus on anti-analysis methods that target dynamic analysis, which security researchers often rely on but are frequently blocked by such defenses. We do not focus on future generation of attack surfaces such as side-channels as such studies has been done by Sihag et al. \cite{sihag2021survey}.

% \subsubsection{Root detection}
% \cite{kellner2019false}

\subhead{Debugger detection}
Android packers employ diverse anti-debugging techniques across both Java and native layers. Common methods include checking debugger status via APIs (e.g., \lstinline|Debug.isDebuggerConnected()|,
\lstinline|getApplicationInfo()|), system properties
(\lstinline|ro.debuggable|), or runtime structures like
\lstinline|gDvm.debuggerConnected|. More advanced defenses hook
components such as \lstinline|JdwpTransport| to block debugger connections altogether~\cite{anti_debugging_android}. Packers also monitor external signs of debugging, such as scanning running processes, checking \texttt{/proc/PID/status} (e.g., \texttt{TracerPid}), inspecting debugger files in \texttt{/data/local/tmp/}, or using the \texttt{ptrace} syscall to detect tracing~\cite{yu2014android}. Timing checks and signal handling further expose debugger interference~\cite{berlato2020jisa}. Together, these layered checks create strong resistance against reverse engineering and dynamic analysis.

\subhead{DBI detection}
Packers defend against \gls{dbi} tools like Frida and Xposed by detecting tampering through memory scans (e.g., injected artifacts or suspicious strings like ``\texttt{frida-agent-32.so}'' in \tt{/proc/self/maps} file)~\cite{suo2025arap}, process and library checks (e.g., \textit{frida-server}, files in \texttt{/data/local/tmp}), or comparing reloaded libraries against in-memory versions to spot hooks. They may also restore trampolined functions (unhooking), use timing checks to detect execution delays, and monitor system-level clues such as thread names (e.g., ``\textit{pool-frida}''), file descriptors, memory permissions (\tt{rwx} which is abnormal), or open communication ports (e.g., 27049). Together, these techniques provide layered defenses that expose hidden \gls{dbi} activity and hinder dynamic analysis~\cite{haupert2018honey}.

\subhead{Systematization of anti-analysis principles}
We present a novel perspective on anti-analysis techniques by organizing them in a way that emphasizes their underlying principles. This approach is particularly beneficial for security researchers seeking to design effective evasion strategies. In contrast to previous systematization~\cite{zhou2022ncscope, ruggia2024unmasking, packdiff}, which primarily focused on grouping detection methods into categories such as emulator detection, root detection, and other specific mechanisms, our taxonomy provides a more \emph{technically actionable} foundation. It is tailored to guide the development of evasion rules that can bypass these techniques in a systematic manner (see Table~\ref{tab:anti-analysis}).
For example, debugger detection can be implemented using either file-based techniques or Java-based APIs. The syntax of the evasion rules can thus correspond to the principles (column ``Anti-analysis principle'') to capture the detection source.

%By identifying the source of information leveraged by the packer to detect anomalies, security researchers can develop tailored evasion strategies. \textcolor{gray}{This source-oriented perspective allows for more systematic and effective bypass mechanisms, as it aligns evasion techniques with the detection vectors used by the packers.}
% put in preamble (once)
% \usepackage{tabularx,makecell}
% \usepackage{booktabs,array}
% \usepackage{url} % or hyperref (already loaded)
\newcolumntype{Y}{>{\raggedright\arraybackslash}X}
\newcommand{\ttpath}[1]{\texttt{\path{#1}}} % breakable monospace paths

% ---- table ----
\begin{table*}[h]
\centering
\scriptsize
\setlength{\tabcolsep}{4pt}
\renewcommand{\arraystretch}{1.15}
\begin{tabularx}{\textwidth}{|c|Y|Y|}
\hline
\textbf{Anti-analysis principle} & \textbf{Detection goal} & \textbf{Example / details} \\
\hline
\multirow{6}{*}{file-based}
  & emulator detection & \ttpath{/dev/qemu}, \ttpath{/proc/cpuinfo}, thermal/sysfs artifacts \\
\cline{2-3}
  & debugger detection & \ttpath{/proc/self/status} (TracerPid), \ttpath{/proc/self/wchan} \\
\cline{2-3}
  & tools detection & \ttpath{/proc/self/maps}, \ttpath{/data/local/tmp/frida-server}, \ttpath{/proc/self/task/TID/stat}, Xposed/JDWP files \\
\cline{2-3}
  & root detection & \ttpath{/system/bin/su}, \texttt{busybox}, RW \texttt{/system} \\
\cline{2-3}
  & Magisk detection & \ttpath{/sbin/.magisk}, \ttpath{/data/adb/magisk} \\
\cline{2-3}
  & repackaging checks & Dex/signature checksum mismatch \\
\hline
\multirow{6}{*}{activity-based}
  & process inspection & \texttt{ps -A}, \texttt{top} for \textit{frida}, \textit{magiskd}, debuggers \\
\cline{2-3}
  & package inspection & Detect apps: Magisk, SuperSU, Xposed, FridaGadget \\
\cline{2-3}
  & emulator detection & \texttt{ro.build.tags}=\textit{test-keys}, abnormal \texttt{ro.product.*} \\
\cline{2-3}
  & dev-mode checks & USB debugging / ADB over TCP enabled \\
\cline{2-3}
  & user-interaction gating & Require gestures or sensor events before running logic \\
\cline{2-3}
  & attestation checks & Play Integrity / SafetyNet results \\
\hline
\multirow{5}{*}{memory-based}
  & tool detection & Abnormal RWX maps, SO injection \\
\cline{2-3}
  & code-integrity checks & Validate function prologues, detect inline hooks \\
\cline{2-3}
  & unpacking resistance & Dex/string/class decryption only in protected regions \\
\cline{2-3}
  & gadget/library probes & Detect \textit{frida-gadget.so} or hooking libs in memory \\
\cline{2-3}
  & ART/loader checks & Hidden class lookups, tampered class loaders \\
\hline
\multirow{3}{*}{timer-based}
  & timing checks & \texttt{System.nanoTime()}, loop delta anomalies \\
\cline{2-3}
  & delayed triggers & Long sleeps, time-of-day based activation \\
\cline{2-3}
  & virtualization timing & I/O/RTT skew typical of emulators \cite{sutter2024dynamic} \\
\hline
\multirow{4}{*}{network-based}
  & TLS pinning & Strict hostname/CA checks, custom trust store \\
\cline{2-3}
  & proxy/MITM detection & Proxy configs, abnormal cert chains \\
\cline{2-3}
  & tool comms detection & Scan JDWP ports, debugger sockets, ADB over TCP \\
\cline{2-3}
  & emulator net probes & Emulator host routes (e.g., 10.0.2.2) \\
\hline
\multirow{5}{*}{Java/Framework}
  & debugger detection & \texttt{Debug.isDebuggerConnected()}, StrictMode \\
\cline{2-3}
  & emulator detection & \texttt{TelephonyManager} IMEI=000000, missing GMS \\
\cline{2-3}
  & code loading guards & Monitor \texttt{DexClassLoader}/\texttt{PathClassLoader} \\
\cline{2-3}
  & runtime API hardening & Checks on clipboard, overlays, sensitive APIs (e.g., \texttt{Class.forName("de.robv.android.xposed.} \\ \texttt{XposedBridge")}) \\
\cline{2-3}
  & reflection checks & Detect suspicious reflective lookups \\
\hline
\multirow{5}{*}{misc / native-level}
  & debugger detection (ptrace) & \texttt{ptrace(PTRACE\_TRACEME)} self/parent attach \\
\cline{2-3}
  & signal tricks & Custom SIGTRAP/SEGV handlers, anti-breakpoint \\
\cline{2-3}
  & environment artifacts & Emulator hardware names (Goldfish/Ranchu), MACs \\
\cline{2-3}
  & sensor probing & Few or missing sensors (NFC, accelerometer, camera) \\
\cline{2-3}
  & early-stage guards & Checks in \texttt{attachBaseContext}/\texttt{JNI\_OnLoad}/\texttt{init\_proc} \\
\hline
\end{tabularx}
\caption{Systematization of Android anti-analysis techniques.}
\label{tab:anti-analysis}
\end{table*}

\subsection{Android Unpackers} \label{sec:unpackers}
In this section, we examine the limitations of existing Android unpackers when confronted with modern packing techniques. Our analysis highlights the growing ineffectiveness of traditional unpacking approaches and motivates a paradigm shift: rather than attempting to fully unpack apps, %embracing a dynamic analysis model, one that 
observing what remains available
alongside packers 
is a more practical and robust strategy. %for analyzing packed Android apps.

\subsubsection{Testing state-of-the-art unpackers} \label{unpackers-tests}
Most modern Android unpackers focus on dumping the decoded Dex code from memory once it becomes available at runtime. These tools generally target the Java layer of an app and rely on either instrumentation of \gls{art} methods or scanning memory regions for Dex content. However, as our experimental evaluation shows, such approaches are increasingly ineffective against modern commercial packers, which use advanced techniques like partial code loading~\cite{packergrind2017leixue}, \gls{jni} interactions, anti-instrumentation (emulation detection~\cite{sahin2018instruction, anti_debugging_android}), and anti-hooking protection to resist analysis. To prevent unpackers from accessing or dumping Dex files in memory, packed apps often hook file-related and memory-related functions to block such operations~\cite{zheng2025gupacker}.

We randomly selected 120 apps packed by Ijiami~\cite{iJiami} as the first group, and another 20 arbitrary packed apps as the second group to evaluate against available unpackers. We do not aim to test our full dataset due to the time-consuming nature of the process. %\textcolor{blue}{Instead, we reference prior studies and highlight that unpacking is often ineffective given the continuous evolution of packers.}

\subhead{ART-based unpackers}
\gls{art}-based unpackers operate by hooking into the functions responsible for loading Dex code, such as \texttt{DexClassLoader} and \texttt{BaseDexClassLoader}. Tools like Youpk~\cite{unpacker_github} and DexHunter~\cite{zhang2015dexhunter} fall in this category. We tested Youpk on the first app group, since the authors claimed Ijiami unpacking support, and found that in 86\% of them, Youpk were not able to dump Dex code even after extending the wait time beyond its recommended duration. Further differential analysis revealed that even in cases where Dex files were dumped, they were incomplete or dependent on native interactions~\cite{happer2021unpacker, Duan2018ThingsYM}. Similarly, DexHunter \cite{zhang2015dexhunter}, which is an academic unpacker designed for older Android versions, failed entirely when tested against our selected apps (second group), largely due to modern packers' use of dynamic loading patterns and \textit{DexLoader} hooking, which disrupts static analysis techniques.

\subhead{Memory-based unpackers}
Memory-based unpackers are also falling short in practice. BlackDex \cite{blackdex_github}, which operates without requiring root or Frida, proved outdated and unable to handle modern Dex release methods, often crashing during execution on the second group of apps. Tools like frida-dexdump~\cite{frida_dexdump_hluwa}, while conceptually powerful, are easily detected by anti-Frida checks, leading to premature termination of the app before dumping can complete. Similarly, KissKiss~\cite{strazzere2014android}, which relies on \texttt{ptrace} to trace the app’s execution, fails due to being detected by anti-debugging techniques early on.

In summary, our results show that existing unpacking tools, whether ART instrumentation-based or memory scanning-based, struggle with the new techniques employed by modern Android packers. %This highlights the need for next-generation methods that can effectively handle encrypted, runtime-loaded code and bypass advanced anti-analysis mechanisms.

\subsubsection{Systematization of Android unpackers}
We have summarized current Android unpackers in Table~\ref{tab:unpacking_solutions}, of which we have tested available ones.
%Since existing unpackers cannot let these packers release all hidden Dex data during unpacking, 
Although unpackers adopt various strategies, ranging from memory scanning and class loading hooks to runtime monitoring, symbolic execution~\cite{xue2021parema}, or hardware assistance~\cite{happer2021unpacker}, their common goal is to recover hidden Dex code from commercial packers.
Their key drawback is that the recovered Dex files do not contain all hidden Dex data~\cite{happer2021unpacker}, as they are not released fully at a given time, if at all (depending on the covered functionality by the duration of interactions). 
%However, they all face major limitations: 
Early unpackers fail due to anti-debugging~\cite{strazzere2014android} and self-modifying code; ART-based approaches are blocked by anti-instrumentation or anti-emulation checks~\cite{Duan2018ThingsYM}; and advanced frameworks struggle with native code, multi-stage releases, or require manual effort. Hardware-assisted and recent solutions~\cite{happer2021unpacker, li2025bpfdex, zheng2025gupacker} remain constrained by heavy instrumentation making them , packer-specific assumptions, and limited coverage (only Dex code) or VM-based protections.

\begin{table*}[h]
\renewcommand{\arraystretch}{1.3}
\centering
\scriptsize
\resizebox{\textwidth}{!}{%
\begin{tabularx}{\textwidth}{|c|X|X|p{2.3cm}|X|X|}
\hline
\textbf{Unpacker} & \textbf{Code coverage} & \textbf{Working platform} & \textbf{Instrumentation level} & \textbf{Available} & \textbf{Year} \\ \hline
KissKiss \cite{strazzere2014android} & Dex & Phone & NA & Yes & 2014 \\ \hline
DexHunter \cite{zhang2015dexhunter} & Dex & Phone (KitKat Android 4.4.3) & DVM \& ART & Yes & 2015 \\ \hline
AppSpear \cite{appspear2015yang} & Dex incomplete & Phone (Android 4.3 and 4.4.2 using AOSP) & DVM & No & 2015 \\ \hline
PackerGrind \cite{packergrind2017leixue} & Dex & Phone (Android 6.0) & Instruction and system level & No* & 2017 \\ \hline
DroidUnpack \cite{Duan2018ThingsYM} & Dex & Emulator (QEMU) & Android framework & No* & 2018 \\ \hline
DexX \cite{dexx2018sun} & Dex & Phone (Nexus 5) & Android kernel and framework & No & 2018 \\ \hline
ReDex \cite{cai2020redex} & Dex & Phone (Android 4.4 DVM, Android 9.0 ART) & Android VM (DVM \& ART, reflection-based) & No & 2020 \\ \hline
Parema \cite{xue2021parema} & Partially Dex codes and VMprotected code & Phone (Android 6.0) & Instruction level (based on PackerGrind) & No* & 2021 \\ \hline
Happer \cite{happer2021unpacker} & Dex & Juno r2 dev board (Android 6.0) & NA & No & 2021 \\ \hline
Gupacker \cite{zheng2025gupacker} & Dex & Phone (Nexus 6) & Android framework & No & 2025 \\ \hline
BPFDex \cite{li2025bpfdex} & Dex & Phone (Pixel 6)\& Emulator & NA & No & 2025 \\ \hline
\end{tabularx}
}
\vspace{0.5em}
\caption{Current Android unpackers sorted by year.  "*": The repositories did not have complete code or were not maintained.}
\label{tab:unpacking_solutions}
\end{table*}

\section{Prevalence Analysis} \label{prevalence-analysis}
We conducted, to the best of our knowledge, the first experimental and large-scale packer prevalence analysis. This is different from previous efforts to examine packed apps, which only involved theoretical analysis or manually checking apps at a small scale. Our purpose is to understand the impact of Android packers on the overall app population which the general public and security practitioners are exposed to.
%a wider spectrum.

We first try to identify packed apps within our dataset and determine the specific packers used. Then, we conduct runtime analysis to assess the effectiveness of anti-Frida and anti-JDB techniques, which are commonly employed to counter \gls{dbi} and Android debugging, respectively. %Unlike previous studies, which provided only partial and often inaccurate analyses, our research offers a more comprehensive evaluation.
\subsection{Dataset}
To conduct a thorough analysis of Android packers and their runtime anti-analysis techniques, we assembled a diverse dataset of 12,341 apps, comprising 7,913 Chinese apps from the 360 App Store~\cite{so_app} and 4,428 non-Chinese apps from APKPure~\cite{apkpure_website}. We separated our dataset into Chinese and non-Chinese apps due to their fundamentally different distribution channels and protection practices. In China, where Google Play is unavailable, developers depend on regional markets that commonly use domestic packers and stricter anti-analysis defenses shaped by local regulations, while non-Chinese apps typically rely on global services focused on intellectual property protection. This distinction also accounts for apps released with region-specific features or protections (see Section~\ref{commercial_packer}), allowing us to capture technical and regulatory differences, highlight ecosystem-specific anti-analysis behaviors, and provide a fairer assessment of packer prevalence worldwide. %\textcolor{red}{it need better wordings. why we separate Chinese and non-Chinese apps in the first place?} \textcolor{purple}{Now take a look?}

\subhead{Downloading process}
The data collection process, carried out between June and August 2024, was fully automated and involved two phases: URL extraction and APK downloading. Selenium~\cite{selenium_python_bindings} was used to scrape download URLs by navigating the interface of both the APKPure and 360 websites, while File Centipede \cite{filecentipede} was employed as a download manager. %Despite APKPure’s strict rate limiting (allowing only 100 downloads per session), we managed to collect a comprehensive set of apps.%
During our analysis, we found 70 apps that were available in both sources. Interestingly, 41 apps shared identical package names but showed different behaviors in our runtime anti-analysis experiments (see Section~\ref{sec:runtime-analysis}), caused by version differences or market-specific modifications. This highlights the importance of testing both regional variants~\cite{guo2025code}.

%\textcolor{purple}{Should I move the categorization problem into the discussion or should I leave it here?}

For app categorization, we relied on a basic mapping approach by translating Chinese category names and matching them with 29 predefined categories used for non-Chinese apps. \textcolor{black}{Although more sophisticated methods, such as those proposed by Alecci et al.~\cite{alecci2024appCategorization}, are recognized as potential future work}, our simplified mapping provided sufficient granularity for the current scope of analysis. Overall, the dataset and collection methodology enabled us to systematically investigate how different types of apps and market contexts affect the use of anti-analysis and packing techniques.

\subsection{Packer Identification}
Commercial packers usually do not try to hide their presence, and their names or other traces are often visible in packed binaries. Tools like APKiD~\cite{apkid} detect such apps by matching known strings, file structures, or libraries. However, APKiD is not maintained as a commercial tool, and its database has outdated entries, which makes it misclassify packers, for example, treating the word “Jiagu” as a specific packer name instead of a general term for protection. This shows the need for better detection tools that consider region-specific factors and are up-to-date, such as NP-Manager~\cite{npmanager}, which specializes in Chinese packers.
%often provides more accurate results.
%prior studies rely mostly on APKiD, overlooking region-specific tools.
While APKiD mainly maps names and strings to Westernized labels, NP-Manager, on the other hand, offers more regionally accurate names and, due to frequent updates, is often more reliable when APKiD fails to detect a packer.

%To validate discrepancies 
To resolve disagreements
between the two tools, we manually inspected the filenames of associated shared libraries (.so) and established mappings between the reported packer names. In certain cases, translations were necessary, or packer labels were associated with broader entities. For example, Baidu~\cite{BaiduApp} is not only a provider of various Internet products, but also offers Android packing services. The following mappings exemplify such findings:
\begin{CJK*}{UTF8}{gbsn}
%\begin{inparaenum}
    SecNeo is equivalent to Bang Bang (Bangcle)~\cite{Bangcle};
    Jiagu (as labeled by APKiD) is associated with 360~\cite{360Dev};
    百度 (Baidu) and 网易易盾 (Yidun) are transliterations of their original Chinese names;
    腾讯御安全{} is recognized as Mobile Tencent Protect;
    Instances labeled as “\textit{UPX modified}” by APKiD were often correctly identified by NP-Manager as Ijiami~\cite{iJiami}.
%\end{inparaenum} 
Given the structural and contextual differences between Chinese and non-Chinese apps, our analysis initially treated these datasets separately. %Comparative insights are presented in the subsequent section.
\end{CJK*}

%\subsubsection{Discussion}
%\subhead{Observation}
\subhead{Findings}
Chinese apps are packed much more frequently than non-Chinese ones. As shown in Figure~\ref{fig:packed_apps_by_cat} and Figure~\ref{fig:packed_apps_by_cat_non-chinese}, the packer identification by APKiD and NP-Manager varies due to differences in their databases and updates. Additionally, categories such as ``finance'' in Chinese apps and "sport and health" in non-Chinese apps contain a higher number of packed apps, indicating the prevalence of packers in specific app categories.

 %Our analysis focused on two key aspects of dynamic instrumentation: \texttt{frida spawn} and \texttt{frida attach} operations. 
%The dataset included 
Our results indicate that 38\% (4,735 of 12,431) of the entire dataset were detected as packed apps. Among the 7,913 Chinese apps, 4,652 (58.8\%) were identified as packed by at least one detection tool (NP-Manager or APKiD), whereas only 83 (2\%) of the 4,428 non-Chinese apps were flagged as packed, supporting the fact that research based solely on non-Chinese apps did not often report failures due to packers. We provide a more detailed breakdown of packer tier distribution across our dataset in Appendix~\ref{packer-tiers}.%These results, summarized in Figure~\ref{fig:packed_apps_by_cat} and~\ref{fig:packed_apps_by_cat_non-chinese}, highlight the significantly higher prevalence of packing in Chinese apps compared to non-Chinese ones.

% Preamble (if not already present):

\pgfplotsset{compat=1.18} % or a suitable version
\usetikzlibrary{patterns}

% In the document
\begin{figure*}[t]
    \centering
    \begin{tikzpicture}
    \begin{axis}[
        ybar,
        bar width=5pt,
        enlarge x limits=0.08,
        ymin=0,
        width=\textwidth,
        height=8cm,
        xlabel={Categories},
        ylabel={\#Packed Apps},
        yscale=0.7,
        symbolic x coords={
            communication, education, finance, life and services, music, 
            parenting, personalization, photography, productivity, reading, 
            shopping, social, sport and health, system, tools, tourism and hotel, 
            travel and location, video
        },
        xtick=data,
        xticklabel style={rotate=25, anchor=east, font=\scriptsize},
        legend style={at={(0.5,1.20)}, anchor=south, legend columns=-1, font=\scriptsize},
        legend image code/.code={%
        \draw[#1,fill] (0cm,-0.1cm) rectangle (0.1cm,0.2cm);},
        nodes near coords,
        every node near coord/.append style={font=\scriptsize, yshift=-1pt, color=black}
    ]

    % Dataset A (NP-Manager)
    \addplot+[ybar, bar shift=-6pt, fill=black!70, draw=black] coordinates {
        (communication, 38) (education, 294) (finance, 387) (life and services, 285)
        (music, 205) (parenting, 290) (personalization, 175) (photography, 320)
        (productivity, 277) (reading, 287) (shopping, 189) (social, 233)
        (sport and health, 253) (system, 208) (tools, 358) (tourism and hotel, 112)
        (travel and location, 150) (video, 236)
    };

    % Dataset B (APKiD)
    \addplot+[ybar, bar shift=6pt, fill=gray!70, draw=black] coordinates {
        (communication, 40) (education, 291) (finance, 384) (life and services, 293)
        (music, 206) (parenting, 341) (personalization, 178) (photography, 320)
        (productivity, 284) (reading, 273) (shopping, 237) (social, 53)
        (sport and health, 269) (system, 213) (tools, 337) (tourism and hotel, 114)
        (travel and location, 267) (video, 230)
    };

    \legend{NP-Manager, APKiD}
    \end{axis}
    \end{tikzpicture}
    \vspace{-10pt}
    \caption[Comparison of Chinese packed apps across categories]{Comparison of Chinese packed apps across different categories using NP-Manager and APKiD.}
    \label{fig:packed_apps_by_cat}
\end{figure*}
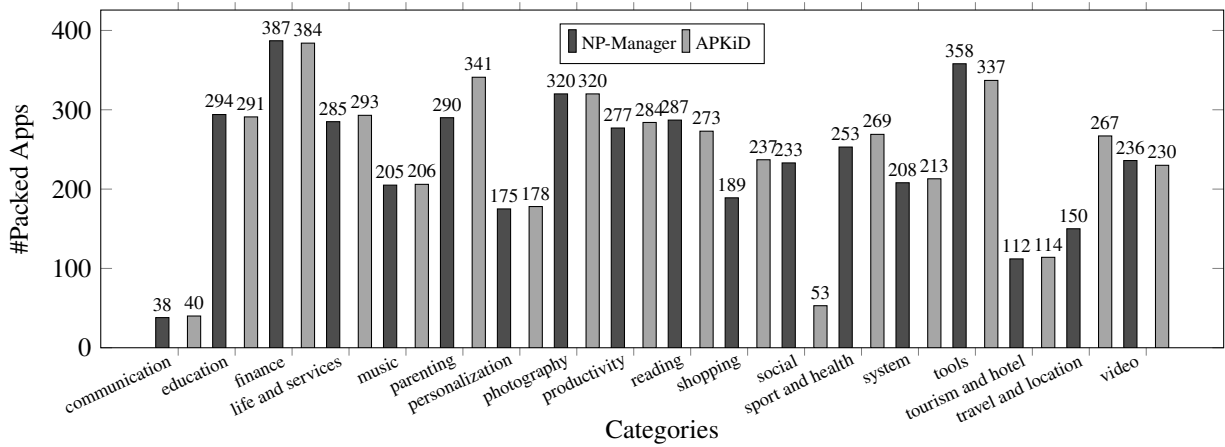

% Required in preamble:
% \usepackage{pgfplots}
% \pgfplotsset{compat=1.18}
% \usetikzlibrary{patterns}

\begin{figure*}[t]
    \centering
    \begin{tikzpicture}
    \begin{axis}[
        ybar,
        bar width=5pt,
        enlarge x limits=0.08,
        ymin=0,
        width=\textwidth,
        height=8cm,
        xlabel={Categories},
        xlabel style={yshift=-4pt},
        ylabel={\#Packed Apps},
        yscale=0.5,
        symbolic x coords={
            business, dating, entertainment, finance, food and drink, 
            house and homes, lifestyle, maps and navigation, 
            medical, news and magazines, parenting, personalization, photography, 
            reading, shopping, social, sport and health, tools, 
            travel and location, vehicle, weather
        },
        xtick=data,
        xticklabel style={rotate=25, anchor=east, font=\scriptsize},
        legend style={at={(0.5,1.55)}, anchor=south, legend columns=-1, font=\scriptsize},
        legend image code/.code={%
        \draw[#1,fill] (0cm,-0.1cm) rectangle (0.1cm,0.2cm);},
        nodes near coords,
        every node near coord/.append style={font=\scriptsize, yshift=-1pt, color=black}
    ]

    % NP-Manager: black fill, white labels
    \addplot+[
        ybar, 
        bar shift=-4pt, 
        fill=black!70, 
        draw=black
    ] coordinates {
        (business, 1) (dating, 2) (entertainment, 0) (finance, 4) (food and drink, 0)
        (house and homes, 0) (lifestyle, 2) (maps and navigation, 1) (medical, 0) (news and magazines, 2)
        (parenting, 0) (personalization, 0) (photography, 1) (reading, 2) (shopping, 2)
        (social, 2) (sport and health, 11) (tools, 4) (travel and location, 3)
        (vehicle, 1) (weather, 0)
    };

    % APKiD: gray fill, black labels
    \addplot+[
        ybar, 
        bar shift=4pt, 
        fill=gray!70, 
        draw=black
    ] coordinates {
        (business, 2) (dating, 1) (entertainment, 1) (finance, 8) (food and drink, 3)
        (house and homes, 3) (lifestyle, 4) (maps and navigation, 1) (medical, 1)
        (news and magazines, 3) (parenting, 1) (personalization, 1) (photography, 1)
        (reading, 2) (shopping, 4) (social, 2) (sport and health, 21)
        (tools, 7) (travel and location, 7) (vehicle, 5) (weather, 1)
    };

    \legend{NP-Manager, APKiD}
    \end{axis}
    \end{tikzpicture}
    \vspace{-10pt}
    \caption[Comparison of non-Chinese packed apps across different categories]{Comparison of non-Chinese packed apps across different categories using NP-Manager and APKiD.}
    \label{fig:packed_apps_by_cat_non-chinese}
\end{figure*}
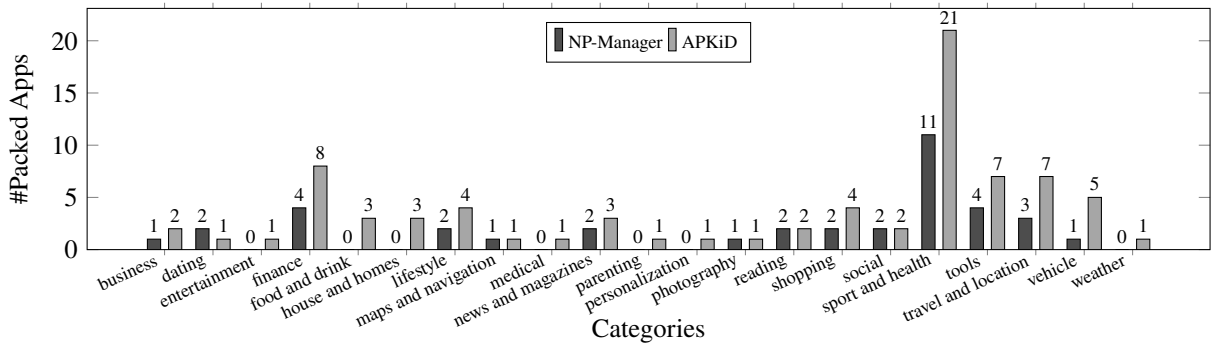

\subsection{Runtime Anti-analysis Techniques} \label{sec:runtime-analysis}
Previous studies have either relied on static analysis 
%for finding anti-analysis tricks in Android apps 
such as matching string patterns, %Besides, dynamic studies identified known anti-analysis principles 
or dynamic analysis, e.g., checking runtime file accesses, which might be inaccurate.
%and assumed the existance of implemented anti-analysis techniques. 
%\textcolor{purple}{Added reference and clarified the reason here}
\textcolor{black}{For example, an app accessing \tt{/proc/self/maps} that can be for loading Dex code during runtime %by packer (benign act) 
\cite{graux2019obfuscated, arp2014drebin} may be mistakenly marked as anti-\gls{dbi} tricks by previous dynamic analysis approaches~\cite{ruggia2024unmasking,druffel2020davinci, packdiff}.}

Instead, we directly test our dataset apps to identify Frida and JDB detection as candidates for \gls{dbi} and Android debugger respectively at runtime. We have used Frida in both ``attach'' and ``spawn'' modes (which can provide more details about initial app interactions and connections) to investigate how common anti-Frida techniques are in practice and not just by heuristics.

Packers react differently when anti-analysis techniques are triggered. Some delay their response, while others react immediately. In some cases, the app crashes; in others, it terminates gracefully or just warns the user that the execution environment is not trusted (e.g., the phone is rooted).
%exhibits altered behavior.

\subsubsection{Frida detection}
%Frida can operate in two modes: ``attach'' (bypassing early-stage detections) or ``spawn''. 
To determine whether an app crashed or terminated due to anti-Frida techniques, we applied multiple crash detection methods in an automated manner.
%\begin{itemize}
First, we %clear logs using \texttt{logcat -v} and then 
monitor the \texttt{logcat} output of each app execution for crash details,
%are typically printed in \texttt{logcat} and stored in tombstone files. We search for strings 
such as "\texttt{FATAL EXCEPTION}" and "\texttt{ANR}" (Application Not Responding), and also verify app liveness using \texttt{ps -A} to detect crashes or terminations. %(\textcolor{red}{How is this automated??} \textcolor{purple}{CHECK its convincing or I have to elaborate}).
Second, we verify the functionality of the Frida agent by:
    \begin{inparaenum}[1)]
        \item executing a simple hooking script to observe expected strings, such as system properties to make sure the Frida agent is still responsive;
        \item examining the Frida client; if it enters the \textit{truncated} mode due to losing communication with the Frida server, we consider this to be an app crash.
    \end{inparaenum}
Third, in the case of app crash or termination, we check the foreground app package name. We use the \texttt{mCurrentFocus} and \texttt{mFocusedApp} properties to identify the focused app and foreground window in Android. If an app was terminated, the foreground app would differ from the one under analysis. Additionally, in some cases, the app request permissions through a dialog box, which our automation tool takes into account.
    
%\end{itemize}
\subhead{Findings}
In total, we found that 2,236 Chinese apps (28.25\%) and 366 non-Chinese apps (8.26\%) deployed anti-Frida mechanisms, targeting either spawn or attach operations (see Table~\ref{tab:frida-attach-spawn-summary}). This corresponds to 2,078 spawn failures and 536 attach failures, which overlap in some apps, resulting in 2,236 unique cases. Among these, 378 Chinese apps (4.77\%) and 159 non-Chinese apps (3.59\%) disrupted Frida in both modes, actively causing crashes or forced terminations. These results indicate that Chinese apps not only use packing more frequently but also integrate stronger runtime defenses.

\begin{table}[H]
    \centering
    \scriptsize
    \setlength{\tabcolsep}{4pt}
    \begin{tabular}{>{\bfseries}lccc}
        \toprule
         & Packed & Non-packed & Total \\
        \midrule
        \multicolumn{4}{c}{\textbf{Chinese apps (7,913 total; Packed $n{=}4{,}652$, Non-Packed $n{=}3{,}261$)}} \\
        \midrule
        Frida spawn failures & 1,574 (33.8\%) & 504 (15.45\%) & 2,078 (26.26\%) \\
        Frida attach failures & 361 (7.8\%) & 175 (5.36\%) & 536 (6.77\%) \\
        \midrule
        \multicolumn{4}{c}{\textbf{Non-Chinese apps (4,428 total; Packed $n{=}83$, Non-packed $n{=}4{,}345$)}} \\
        \midrule
        Frida spawn failures & 28 (33.7\%) & 234 (5.38\%) & 262 (5.91\%) \\
        Frida attach failures & 23 (27.7\%) & 240 (5.52\%) & 263 (5.93\%) \\
        \bottomrule
    \end{tabular}
    \caption[Frida spawn/attach failures in Chinese vs. non-Chinese apps]{
    Frida spawn and attach failure statistics across Chinese and non-
Chinese apps. Percentages are per subgroup.
    }
    \label{tab:frida-attach-spawn-summary}
\end{table}

\subsubsection{JDB detection}
To analyze anti-debugging techniques at runtime, the app must be executed in a debuggable mode, allowing it to launch with the \textit{Waiting for debugger} option. This typically requires setting the \texttt{android:debuggable} flag in the \texttt{AndroidManifest.xml} file of the APK. However, modifying this flag may trigger anti-tampering mechanisms~\cite{notrepackage}, thereby compromising the integrity of the analysis. To overcome this limitation, we adopted an approach based on modifying the Android source code to enable debugging for non-debuggable apps. This modification was tested on a Pixel device using the Youpk project~\cite{unpacker_github}.

The dynamic analysis workflow involves forwarding the JDWP port, starting the target app in debug mode using \texttt{am start -D -n ACTIVITY\_NAME}, and attaching to it via JDB. If the app requests runtime permissions, they are automatically granted to preserve the fidelity of the analysis. Any unexpected app crash or premature termination of the JDB session is interpreted as an indication that an anti-debugging mechanism has been triggered. This methodology enables a comprehensive examination of runtime anti-debugging techniques, including those
%that have traditionally been 
studied through static analysis.

% It is important to note that our findings represent a conservative lower bound. Some anti-analysis techniques may have been triggered beyond our defined testing conditions. In particular, the runtime analysis for each app was limited to 15 seconds, which may have excluded detection routines that activate after longer execution periods. %Consequently, our results may not capture all methods employed by packed apps to detect JDB-based debugging. 

\subhead{Findings} Our results shows that 1,960 apps crashed while attaching to \gls{jdb} in total. Interestingly, the number of non-Chinese non-packed apps that experienced crashes was higher than that of the non-Chinese packed apps. Our findings are summarized in Table~\ref{tab:jdb-summary}.

\begin{table}[H]
    \centering
    \scriptsize
    \setlength{\tabcolsep}{4pt}
    \renewcommand{\arraystretch}{1.1}
    \begin{tabular}{>{\bfseries}lccc}
        \toprule
         & Packed & Non-packed & Total \\
        \midrule
        \multicolumn{4}{c}{\textbf{Chinese apps (packed $n{=}4{,}652$, non-packed $n{=}3{,}261$), 7,913 total}} \\
        \midrule
        JDB failures & 872 (18.74\%) & 591 (18.12\%) & 1,463 (18.48\%) \\
        \midrule
        \multicolumn{4}{c}{\textbf{non-Chinese apps (packed $n{=}83$, non-packed $n{=}4{,}345$), 4,428 total}} \\
        \midrule
        JDB failures & 11 (13.25\%) & 486 (11.18\%) & 497 (11.22\%) \\
        \bottomrule
    \end{tabular}
    \caption[JDB failure rate in Chinese vs. non-Chinese apps]{
    JDB failure statistics across Chinese and non-Chinese apps. 
    Percentages are per subgroup.
    %Percentages are calculated within each subgroup (e.g., 872 packed failures out of 4,652 Chinese packed apps = 18.74\%).
    }
    \label{tab:jdb-summary}
\end{table}

\section{Living with the Packers}

%Since unpacking is not effective for dynamic analysis, we suggest hiding the presence of dynamic analysis tools by evading anti-analysis techniques and allowing the dynamic analysis tools to live alongside the packer. %These techniques involve checks performed by packers to detect abnormalities in the process context or environment that are introduced by dynamic analysis tools.
In this section, we implement our proposed idea of living with packers, given the ineffectiveness of unpacking, 
%By ``enabling dynamic analysis'', we mean bypassing such 
by bypassing runtime checks that would otherwise cause the app to crash or terminate itself and still allowing the analysis tool to access whatever remains available at runtime (already released/extracted for execution).

\subhead{Balancing stealth and usability}
%\textcolor{red}{I think this is not the right place for this subsection}
Frida provides a flexible and programmable interface widely adopted by security analysts, but its popularity has led to the development of numerous detection methods. In contrast, eBPF operates at the kernel level, offering lower visibility to the app and making detection harder. However, eBPF lacks Frida’s high-level programmability (e.g., Javascript and Python) and usability for Android app analysis (e.g., \url{https://codeshare.frida.re}). Thus, we choose to use one to enable the other to achieve a balance between stealth and practicality to support effective yet undetectable dynamic analysis.

% \input{tables/ebpf-compare-other-hooking-tbl}

% Required in preamble:

Unlike prior dynamic analysis approaches that rely on AOSP/kernel modifications or emulation, leading to poor portability, instability and maintenance overhead, {\ourframework} remains lightweight, scalable, and safe while enable existing DBIs like Frida. {\ourframework} achieves practical analysis without app modification, supports Google Play services, and offers a balanced trade-off between stealth, portability, and effectiveness (see Table~\ref{tab:comparison-instrumentation}).

% Required in preamble:

\begin{table*}[!t]
    \centering
    \scriptsize
    \renewcommand{\arraystretch}{1.2}
    \setlength{\tabcolsep}{5pt}
    \begin{tabular}{|l|c|c|c|c|c|c|c|}
        \hline
        & \textbf{No App Modification} & \textbf{Not Visible to App} & \textbf{Portable} & \textbf{Scalable} & \textbf{Safe} & \textbf{System-Level Monitoring} & \textbf{Google Play Services} \\
        \hline
        AOSP Modification   & \cmark & \cmark & \xmark & \cmark & \xmark & \cmark & \xmark \\
        Kernel Module       & \cmark & \cmark & \xmark & \cmark & \xmark & \cmark & \xmark \\
        App Modification    & \xmark & \xmark & \cmark & \xmark & \xmark & \xmark & \cmark \\
        Library Injection   & \cmark & \xmark & \cmark & \cmark & \cmark & \xmark & \cmark \\
        \textbf{\ourframework} & \cmark & \cmark & \cmark & \cmark & \cmark & \cmark & \cmark \\
        \hline
    \end{tabular}
    \caption[Comparison of instrumentation techniques in Android app analysis]{Comparison of instrumentation techniques for enabling Android app dynamic analysis.}
    \label{tab:comparison-instrumentation}
\end{table*}

\subsection{{\ourframework} Overview}
We propose {\ourframework}, an eBPF-based evasion engine that is able to bypass anti-analysis techniques at runtime without being detected. It accepts \gls{der}s as configuration files and applies evasion rules on the fly to enable dynamic analysis tools such as Frida. To the best of our knowledge, {\ourframework} is the first offensive eBPF application that manipulates app memory to evade detection. Previous eBPF-based works have mostly focused on monitoring and instrumentation \cite{zhou2022ncscope}.

\begin{figure}[h]
    \centering
    \hspace*{-0.5cm} % adjust this value
    \includegraphics[width=1\columnwidth]{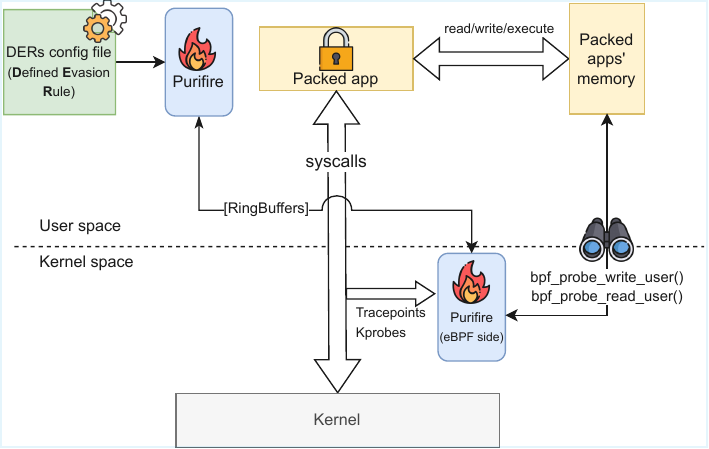}
    \caption[Purifire overview.] {Purifire Overview. This diagram illustrates how {\ourframework} captures system calls and manipulates application memory using eBPF.}
    \label{fig:purifire}
\end{figure}

Figure~\ref{fig:purifire} provides an overview of {\ourframework}. {\ourframework} consists of three components: a userland program, an eBPF kernel-side program, and a configuration file named \gls{der}. {\ourframework} applies runtime memory patches based on \gls{der}s to the app’s process context in order to bypass anti-analysis techniques.

\subsection{Architecture}
%The design of \ourframework{} is driven by the need to bypass runtime anti-analysis techniques in packed Android apps without leaving detectable artifacts in user space. 
Our approach combines a configuration-driven controller in user-space with an eBPF-based enforcement engine in kernel space, providing stealthy, flexible, and adaptive evasion at runtime. The kernel-space and user-space components of \ourframework{} achieve two purposes explained in the following. %complementary modules for packed app analysis.

\subhead{Assisted analysis} It provides syscall-level observation for the \gls{der} authors to find suspicious candidate to manipulate. We have also developed a syscall-memory map generator that finds which memory regions have been allocated and what syscalls are originally generated from what memory regions. We trace \texttt{mmap} and \texttt{mprotect} syscalls to generate memory maps, including those originating from the packer library at runtime. Each syscall is then linked to its corresponding memory region to determine its caller function. Combined with stack backtraces obtained using the \tt{bpf\_get\_stack()} eBPF helper, this process help \gls{der} authors identify suspicious syscalls more accurately and avoid manipulating unrelated regions during the evasion phase (see Appendix~\ref{appI}).

\subhead{Runtime evasion} It filters syscall events and manipulates memory when specific conditions defined in the \gls{der}s are met, such as a particular syscall argument or a specific value in a memory region.

\subhead{Defined evasion rules} \gls{der}s are {\ourframework}'s configuration files that define evasion strategies. Each \gls{der} rule consists of two parts: condition and evasion. The \textit{condition} specifies filters to exclude irrelevant events, ensuring that {\ourframework} only targets anti-analysis techniques and avoids manipulating the wrong locations. Listing~\ref{lst:der_examples} illustrates how a \gls{der} can focus on a specific syscall (event) associated with the anti-analysis protection of a packer library. %(see Section~\ref{workflow} \gls{der} design). 
For example, it can filter based on the package name (comm) and then select the syscall to manipulate. The \textit{evasion} part of the rule identifies the userspace pointer that {\ourframework} intends to modify by writing directly into the app’s memory.

\begin{figure}[t]
\centering
\lstset{
  basicstyle=\ttfamily\tiny,
  numbers=none,
  frame=single,
  showstringspaces=false,
  breaklines=true
}
\begin{lstlisting}[language=json, caption={DER configuration examples.}, label={lst:der_examples}]
/* Example 1 */
{
  "condition": {
    "comm": "com.example.test",
    "tname": "*",
    "syscall": "openat",
    "args": { "1": "/proc/self/task/" }
  },
  "evasion": {
    "where": "args1",
    "data": "/data/local/tmp/fake"
  }
}

/* Example 2 */
{
  "condition": {
    "comm": "com.example.test",
    "tname": "*",
    "syscall": "mprotect",
    "args": { "0": "arg0", "1": "0xde5c0", "2": "0x5" },
    "data": "\\x28\\x10\\x80\\xd2\\x01\\x00\\x00\\xd4"
  },
  "evasion": {
    "where": "args0 + 0x6aae4",
    "data": "\\x00\\x00\\x80\\xd2\\x1f\\x20\\x03\\xd5"
  }
}
\end{lstlisting}
\vspace{-20pt}
\end{figure}

%\subsubsection{Architecture}
%The evasion module works as follows:
\subhead{Components}
%\begin{enumerate}
%    \item \textbf{User-space (controller):}  
    The user-space program (controller) is responsible for parsing \gls{der} configuration files and translating them into filtering and patching policies. It loads these rules into the kernel via eBPF interfaces e.g., \textit{RingBuffers}.
%    \item \textbf{Kernel-space (eBPF Engine):}  
    On the kernel-space side, the core enforcement logic is implemented as eBPF programs attached to kprobes and tracepoints for critical system calls (e.g., \texttt{ptrace}, \texttt{mprotect}, \texttt{openat}, \tt{readlinkat} and \texttt{prctl}). They monitor syscall arguments, thread identifiers, and calling context. When a \gls{der} matches (via \texttt{bpf\_probe\_read\_user()}), the eBPF programs can alter the syscall outcome or patch user memory directly via \texttt{bpf\_probe\_write\_user()}.
 %   \item \textbf{Target App and Memory:}  
    The packed app continues executing as normal, with its runtime environment transparently modified. %Anti-analysis checks that would otherwise cause app's termination or crash.
%\end{enumerate}

% \input{figures/DER_example_json}

\subsection{Workflow} \label{workflow}
\begin{figure}[h]
    \centering
    \hspace*{0.09cm}
    \vspace{-15pt}
    \includegraphics[width=1\columnwidth]{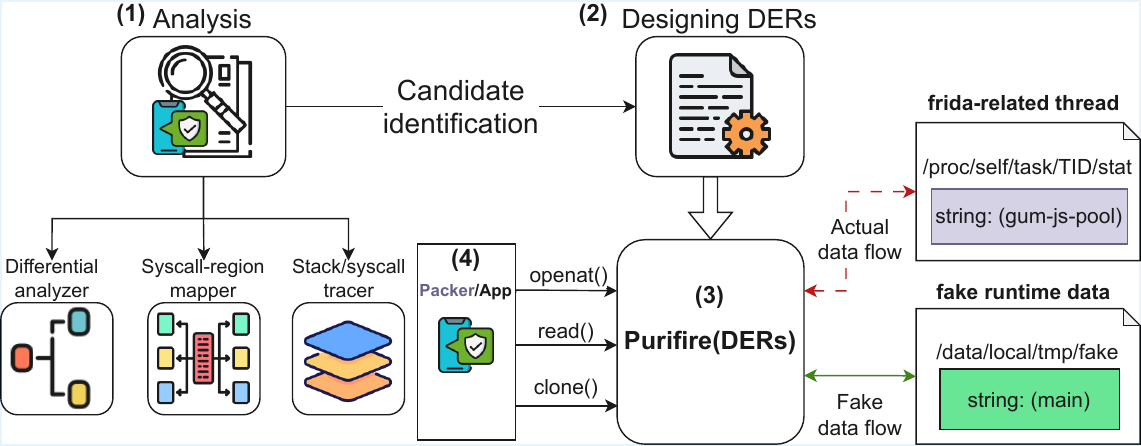}
    \caption[Purifire workflow.] {\ourframework{} example scenario. This diagram illustrates how \ourframework{} uses \gls{der}s to bypass runtime anti-analysis techniques triggered by syscall events.}
    \label{fig:workflow}
\end{figure}

As one-time effort per packer version, a \gls{der} author can, manually, with the help of \ourframework{}'s assisted analysis, identify candidate syscalls to manipulate and refine them by targeting specific arguments. The author must then map these syscalls to the corresponding anti-analysis principles (see Table~\ref{tab:anti-analysis}) that cause the app to crash and construct \gls{der}s to bypass them. Figure~\ref{fig:workflow} illustrates how the \gls{der} example in Listing~\ref{lst:der_examples} can be applied to evade an anti-Frida technique from the file-based tool detection category in Table~\ref{tab:anti-analysis}. As demonstrated, a single anti-analysis method may expose multiple manipulation points (e.g., \tt{openat()}, \tt{read()}, \tt{clone} syscalls) that the \gls{der} author can exploit to construct different \gls{der}s. In some cases, it is also possible to completely nullify the anti-analysis protection by preventing the creation of its monitoring thread (patching code that calls \tt{clone()}).

% DERs are not targeting syscall argument manipulation only but it can also overwrite the syscall results (stored data).

Considering the workflow of packers, multi-stage techniques~\cite{Duan2018ThingsYM}, dynamic code loading (see Figure~\ref{fig:packer-stages} in Appendix~\ref{sec:stages}), and data encryption, the event-driven evasion engine is more effective than fixed patches, which are case-specific and easily broken by minor packer changes.

\subhead{Runtime memory patching} From the perspective of evasion, if the code and data can be patched statically, meaning the anti-analysis logic already exists in the APK and does not require runtime decryption, it is possible to patch the APK directly but when it comes to modern packers that decrypt code and data at runtime, any beforehand patching becomes challenging. As a result, runtime memory patching is a more promising approach to bypass anti-analysis techniques.

\subhead{DER considerations}
Designing \gls{der}s requires experimenting with different data manipulation strategies to find a working bypass, which is often challenging and require manual analysis in this cat-and-mouse setting. Manipulation points can vary; for example, altering an \texttt{openat()} syscall argument to redirect the app to a dummy file, or modifying the subsequent \texttt{read()} syscall for finer-grained control. In general, the fewer modifications made to the app, the better stealthiness can be ensured.
%of the bypass regarding integrity checks of the packers.

The DER author can observe which syscall causes the app to crash or terminate and perform differential analysis by comparing the app’s behavior when run with Frida versus without it, stack trace analysis, and syscall-region mapping (assisted analysis using \ourframework{} Figure~\ref{fig:workflow}). %This differential analysis, provided by our analysis module, 
This process can highlight candidate syscalls for investigation, narrowing the search space for effective evasion rules.

Anti-analysis techniques are not limited to syscalls and may also rely on memory checks, such as \texttt{strstr()} or inline memory comparisons, which cannot be traced via syscalls. In such cases, a \gls{der} author can leverage stealthy eBPF-based memory dumpers~\cite{lemon_eurecom_s3} to locate suspicious strings in the app’s memory and calculate offsets for use in the \gls{der}. For example, after decryption, the Ijiami packer actively opens \texttt{/proc/self/wchan} based on syscall origin maps (see Appendix~\ref{appI}), while other checks can be inferred from previously dumped data, such as locating the \texttt{ptrace\_stop} string offset (part of debugging fingerprints) and patching it once loaded into memory. A final \gls{der} can then be defined by filtering \texttt{mprotect} or \texttt{mmap} syscalls and validating arguments such as size (in Listing~\ref{lst:der_examples} Example 2, \tt{0xde5ce} is the size) and expected data. If these conditions are met, \ourframework{} overwrites the memory location (in Listing~\ref{lst:der_examples} Example 2, \tt{args0} as base address and \tt{0x6aae4} as code offset resulted from stack trace and memory-syscall mapping from previous runs). Another example is identifying suicidal code (instructions that cause the app to crash or terminate) and patching it. With the stack trace provided by \ourframework{}'s assisted analysis, the \gls{der} author can locate caller functions and patch the \texttt{kill} syscall responsible for termination. Even if the packer marks this code as read-and-execute only, the \gls{der} author can still calculate the syscall offset and apply the patch on \tt{mprotect} event (see Listing~\ref{lst:der_examples}).

%\textcolor{purple}{the DER community based and maintenance addressed here}

In summary, {\ourframework}'s assisted analysis tracks memory region permission changes and provides stack trace–based offsets to help \gls{der} authors design precise filters and patches. Despite the (semi-)manual effort, the resulting DER configuration files can then be shared and maintained by the community, enabling other analysts to reuse them for dynamic analysis and contributing to a growing public knowledge base. % (see Listing~\ref{lst:der_examples}).

\iffalse
\subsubsection{Novelty and Strengths}
{\ourframework} introduces several unique features compared to prior approaches:
\begin{itemize}
    \item \textbf{First use of eBPF for active evasion:} While prior work used eBPF for monitoring, we employ it to actively disable anti-analysis tricks at runtime.
    \item \textbf{Stealth by design:} Packers are not able to detect {\ourframework} as it works in kernel-level.
    \item \textbf{Rule-driven flexibility:} Unlike static patching, the \gls{der} mechanism enables flexible and fine-grained evasion strategies tailored to different apps.
    \item \textbf{Complementary to existing tools:} Purifire does not attempt unpacking or reengineering but rather ensures that existing dynamic analysis frameworks (e.g., Frida, Xposed) can operate undisturbed.
\end{itemize}
\fi

\section{Evaluation} \label{evaluations}
%Despite the {\ourframework}'s innovation in bypassing anti-analysis tricks we also demonstrate its effectiveness that {\ourframework} is able to enhance the results of other works.
In this section, we demonstrate experimentally and comparatively how \ourframework{} can improve the outcome of state-of-the-art Android app security analysis, e.g., academic papers involving hooking that could have been affected by packers (see Section~\ref{frida-usecase}). This shows how \ourframework{} can benefit security research, e.g., against undesired app behavior.
Note that certain works did not report how packed apps affected their results or a dataset without Chinese apps was used.

\subhead{Environment setup}
We conducted our experiments on a Pixel 7 Pro running Android 15 with kernel version 5.10. The device was rooted using Magisk~\cite{magisk_github} (version 28.1) with Zygisk enabled. To facilitate dynamic analysis, we installed the Frida server and deployed our tool, {\ourframework}, alongside the \gls{der}s to the \texttt{/data/local/tmp/purifire/} directory.

%\textcolor{red}{Where's our original evaluation??? We still need them, supplemented by the new ones below.}

\subsection{Frida with \ourframework{}} \label{sec:first_eval}
Our prevalence analysis (see Section~\ref{prevalence-analysis}) confirmed that 2,340 apps in our dataset employed anti-analysis techniques that lead to crash/termination while being spawned under Frida instrumentation. We repeated the detection phase with \ourframework{} enabled and configured \gls{der}s to bypass known tricks, including anti-root 
%(hiding the presence of the \texttt{su} binary) 
and anti-Frida (covering artifacts described in Section~\ref{anti-analysis-tricks}). %\textcolor{purple}{HERE I don't like 5.30\% to be shown but you said otherwise they assume we are biased} As a result, 
We successfully ran Frida on additional 662 apps (enabled Frida on 28.2\% of those with anti-analysis tricks or 5.30\% on our entire dataset) that would otherwise crash or terminate without \ourframework{}. Note that the 662 apps were what our DERs enabled us to cover, not all the possibilities, e.g., if DERs for more packers were provided, the corresponding number of more apps would be saved.

Table~\ref{tab:frida-evaluation-packed} presents the apps on which \ourframework{} successfully enabled Frida, highlighting clear differences between packed and non-packed categories: packed apps are concentrated in tools, finance, and system, whereas non-packed apps are more prevalent in social, reading, and photography. Table~\ref{tab:frida-evalusations-packers} further illustrates the packer distribution in our evaluation, showing that Jiagu, SecNeo, and yidun together account for nearly three quarters of packed apps, underscoring both \ourframework{}’s effectiveness and the dominance of a few commercial solutions in Android app protection.

\begin{table}[H]
\centering
\caption{\ourframework{}-enabled apps (packed vs. non-packed).}
\fontsize{4pt}{6pt}\selectfont
\renewcommand{\arraystretch}{0.8}
\setlength{\tabcolsep}{3pt}
\resizebox{\linewidth}{!}{%
\begin{tabular}{|l|c|c|c|c|}
\hline
\multirow{2}{*}{\textbf{Category}} & \multicolumn{2}{c|}{\textbf{Packed apps (N=448)}} & \multicolumn{2}{c|}{\textbf{Non-packed apps (N=214)}} \\ \cline{2-5}
 & \#bypassed & percentage (\%) & \#bypassed & percentage (\%) \\ \hline
communication      & 3  & 0.67\%  &  9 & 4.21\% \\ 
education          & 38 & 8.48\%  &  8 & 3.74\% \\ 
finance            & 48 & 10.71\% & 17 & 7.94\% \\ 
life\_services     & 26 & 5.80\%  & 12 & 5.61\% \\ 
music              & 19 & 4.24\%  &  6 & 2.80\% \\ 
parenting          & 18 & 4.02\%  &  3 & 1.40\% \\ 
personalization    & 15 & 3.35\%  &  5 & 2.34\% \\ 
photography        & 29 & 6.47\%  & 16 & 7.48\% \\ 
productivity       & 25 & 5.58\%  & 10 & 4.67\% \\ 
reading            & 30 & 6.70\%  & 18 & 8.41\% \\ 
shopping           & 18 & 4.02\%  &  9 & 4.21\% \\ 
social             & 28 & 6.25\%  & 18 & 8.41\% \\ 
sport \& health    & 21 & 4.69\%  & 12 & 5.61\% \\ 
system             & 30 & 6.70\%  &  5 & 2.34\% \\ 
tools              & 55 & 12.28\% &  7 & 3.27\% \\ 
tourism\_hotel     &  8 & 1.79\%  &  1 & 0.47\% \\ 
travel\_location   & 23 & 5.13\%  &  8 & 3.74\% \\ 
video              & 14 & 3.12\%  & 12 & 5.61\% \\ \hline
\textbf{Total}     & 448 & 100\%  & 214 & 100\% \\ \hline
\end{tabular}%
\label{tab:frida-evaluation-packed}
}
\end{table}

\begin{table}[H]
\centering
\caption{\ourframework{}-enabled packed apps by packer names (N=448).}
\fontsize{4pt}{6pt}\selectfont
\renewcommand{\arraystretch}{0.8}
\setlength{\tabcolsep}{5pt}
\resizebox{\linewidth}{!}{%
\begin{tabular}{|l|c|c|}
\hline
\textbf{Packer} & \textbf{\#Bypassed apps} & \textbf{Percentage (\%)} \\ \hline
Jiagu            & 161 & 35.94\% \\ 
SecNeo.A         &  99 & 22.10\% \\ 
yidun   &  40 &  8.93\% \\ 
SecNeo.B         &  36 &  8.04\% \\ 
Ijiami   &  22 &  4.91\% \\ 
Tencent Protect  &  20 &  4.46\% \\ 
SecNeo  &  14 &  3.13\% \\ 
Baidu    &   6 &  1.34\% \\ 
UPX / sharelib   &   6 &  1.34\% \\ 
CrazyDog Wrapper &   3 &  0.67\% \\ 
Bangcle  &   2 &  0.45\% \\ 
Approov          &   1 &  0.22\% \\ \hline
\textbf{Total}   & 448 & 100\% \\ \hline
\end{tabular}
\label{tab:frida-evalusations-packers}
}
\end{table}

\subsection{Device Fingerprinting with \ourframework{}}
Heid et al.~\cite{heid2024haven} used Frida to hook specific \gls{api}s in order to identify apps that perform device fingerprinting. We evaluated their tool on our dataset and measured the number of \gls{tdf} it produced. We then repeated the experiment, this time running {\ourframework} alongside their tool, and collected the resulting data. The number of identified \gls{udf} increased significantly when {\ourframework} was used. We ran their tool on our entire dataset.
\begin{itemize}
\setlength{\itemsep}{-5pt}
    \item Number of \gls{udf} without {\ourframework}: 79,260
    \item Number of \gls{udf} with {\ourframework}: 131,173
\end{itemize}
The number of \gls{udf}s still observed without \ourframework{} is due to anti-analysis checks triggered with a delay, during which the app or packer continued collecting \gls{udf}s. Table~\ref{tab:udf_summary} shows that Purifire increased \gls{udf} capture by 65.5\% across 1,214 apps. The largest gains appeared in music (+555.4\%), beauty (+273.6\%), event (+184.4\%), personalization (+168.3\%), and news\_and\_magazines (+173.8\%), while categories like video (+1.7\%) and art (+2.5\%) saw minimal change. These results highlight \ourframework{}’s effectiveness in enhancing fingerprint visibility, particularly in user-facing domains where tracking and personalization are common.

\begin{table}[H]
\centering
\caption{Numbers of UDF/TDF captured before/after applying \ourframework{} by category. UDF is from deduplicating TDF.}
\label{tab:udf_summary}
\fontsize{4pt}{6pt}\selectfont
\renewcommand{\arraystretch}{0.8}
\setlength{\tabcolsep}{3pt}
\resizebox{\linewidth}{!}{%
\begin{tabular}{|l|r|r|r|r|r|r|}
\hline
\textbf{Category} & \textbf{\#Apps} & \textbf{$\sum$UDF$_{before}$} & \textbf{$\sum$UDF$_{after}$} & \textbf{$\sum$TDF$_{before}$} & \textbf{$\sum$TDF$_{after}$} & \textbf{UDF $\Delta$ (\%)} \\ \hline
art                & 12  & 1688  & 1730  & 26490  & 27609  & 2.49\% \\
beauty             & 21  & 769   & 2873  & 10085  & 40091  & 273.60\% \\
business           & 4   & 549   & 566   & 7369   & 8616   & 3.10\% \\
communication      & 32  & 1612  & 3884  & 29385  & 67273  & 140.94\% \\
dating             & 45  & 4984  & 6101  & 65284  & 88356  & 22.41\% \\
education          & 45  & 2191  & 4327  & 33776  & 87380  & 97.49\% \\
entertainment      & 9   & 663   & 884   & 9438   & 18237  & 33.33\% \\
event              & 21  & 920   & 2616  & 16274  & 43666  & 184.35\% \\
finance            & 43  & 3694  & 5517  & 54726  & 83184  & 49.35\% \\
food\_and\_drink   & 65  & 4610  & 8291  & 67783  & 135895 & 79.85\% \\
house\_and\_homes  & 32  & 3562  & 4208  & 51070  & 64040  & 18.14\% \\
life\_services     & 11  & 806   & 976   & 13930  & 16323  & 21.09\% \\
lifestyle          & 28  & 2902  & 3311  & 46206  & 56885  & 14.09\% \\
location           & 1   & 171   & 178   & 1492   & 1971   & 4.09\% \\
maps\_and\_navigation & 45 & 4011 & 5635  & 66324  & 98694  & 40.49\% \\
medical            & 23  & 2039  & 2474  & 35117  & 44783  & 21.33\% \\
music              & 192 & 1889  & 12381 & 23090  & 267838 & 555.43\% \\
news\_and\_magazines & 97 & 4319 & 11824 & 45764  & 209824 & 173.77\% \\
parenting          & 13  & 1488  & 1561  & 27159  & 28118  & 4.91\% \\
personalization    & 32  & 1188  & 3187  & 22750  & 64583  & 168.27\% \\
photography        & 31  & 3412  & 3934  & 56701  & 66690  & 15.30\% \\
productivity       & 10  & 481   & 678   & 12185  & 15825  & 40.96\% \\
reading            & 34  & 3238  & 3682  & 55369  & 64524  & 13.71\% \\
shopping           & 58  & 4644  & 7182  & 64207  & 109206 & 54.65\% \\
social             & 78  & 3690  & 7086  & 56713  & 118919 & 92.03\% \\
sporthealth        & 65  & 6315  & 8388  & 81439  & 125831 & 32.83\% \\
system             & 30  & 1552  & 2204  & 23969  & 49773  & 42.01\% \\
tools              & 29  & 2875  & 3667  & 41799  & 59668  & 27.55\% \\
tourism\_hotel     & 3   & 80    & 169   & 2388   & 3490   & 111.25\% \\
travel\_location   & 50  & 3711  & 5184  & 65612  & 101753 & 39.69\% \\
vehicle            & 9   & 803   & 1002  & 10431  & 15408  & 24.78\% \\
video              & 12  & 1383  & 1407  & 22575  & 23675  & 1.74\% \\
weather            & 34  & 3021  & 4066  & 52758  & 74493  & 34.59\% \\ \hline
\textbf{Total}     & 1,214 & 79,260 & 131,173 & 1,199,658 & 2,282,621 & 65.50\% \\ \hline
\end{tabular}%
}
\end{table}

\subsection{Covert Identifier Detection with \ourframework{}}
Dong et al.~\cite{dong2024exploring} used Frida in their dynamic analysis pipeline to collect third-party SDK file operations on external storage that allow them to track users across multiple apps, as a covert channel. They acknowledged failures with a subset of apps due to packing services. In consideration of this, unlike our previous two experiments, we opted to be more targeted to packed apps (more accurately apps with anti-analysis techniques that hindered the authors' analysis). 

To measure \ourframework{}’s benefit, we randomly selected 70 apps from the 662 apps which we have the DERs for from Section~\ref{sec:first_eval}. This small subset is due to the nature of Dong et al.'s analysis requiring manual interaction with each app's UI, hence limiting the scale of the test. We observed that on 35 such apps, \ourframework{} boosted the total number of observed shared-location file accesses from 267 to 7,565 and unique accesses from 133 to 3,355, indicating an improvement of 27× more total and 24× more unique file operations for tracking users.

% \subsection{SSL Pinning} \textcolor{purple}{TBD}

\section{Discussion}

\subhead{Frida and \ourframework{} cooperation}
{\ourframework} operates independently of Frida, allowing both tools to function alongside each other since they target different layers of the system. This enables security practitioners to leverage {\ourframework} in conjunction with Frida scripts to effectively bypass anti-analysis checks while maintaining dynamic instrumentation capabilities.

\subhead{Low deployment requirement}
To use {\ourframework}, merely a rooted phone is needed. Obviously, this can trigger root detection mechanisms, which are then bypassed with well-defined DERs. In comparison, most unpackers are more invasive, requiring, e.g., custom AOSP builds~\cite{appspear2015yang,dexx2018sun}, ART modifications~\cite{zhang2015dexhunter,zheng2025gupacker}, loading kernel modules or even hardware support~\cite{happer2021unpacker}.

\subhead{Code and data integrity checks by packers}
Currently, \ourframework{} targets \texttt{syscall\_enter} events, and it is possible to roll back changes and memory patches at \texttt{syscall\_exit} by reverting the \gls{der} configurations. This helps avoid integrity checks performed by packers at runtime.

% \textcolor{purple}{Did I mentioned this enough times? :))}

% \subhead{Obfuscation impacts}
% Dumped Dex code alone mostly not enough for analyzing app behaviors as they need to be analyzed at runtime. However, both Java and  native code obfuscations are  Some non-academic works suggested \textit{de-shelling} which consist of putting dumped Dex code and other app's resources back into an APK file, installing it and analyze it at runtime, this time without packer presence.

% \subhead{Nothing to hook}
% App analyst may 

\subhead{How \ourframework{} helps security researchers}
By enabling tools like Frida on packed apps, Purifire revives research areas such as privacy leakage, SDK misuse, and fingerprinting studies that were previously hindered. Through bypassing runtime anti-analysis techniques, it allows Frida-based inspection for network traffic analysis (e.g., detecting \gls{pii} exposures, certificate pinning, and TLS/SSL issues), memory analysis (e.g., heap dumping), and instrumentation tasks such as function hooking~\cite{sutter2024dynamic}.

\section{Limitations} \label{sec:limit}

\subhead{DER creation}
\ourframework{} is an evasion engine that relies on DERs (the rules) to function. While this involves manual effort and domain knowledge (although with the help of \ourframework{}'s assisted analysis), the effort is one-time per packer version and can be circulated in the community. %\textcolor{purple}{Incorrect \gls{der}s may overwrite memory regions unrelated to anti-analysis tricks, potentially causing app crashes.}

% \subhead{memory only anti-analysis tricks}{\ourframework} can only bypass syscall- or event-based techniques and cannot handle purely memory-based anti-analysis methods. However, in future work, with proper \gls{der} memory/value-based rules, these techniques could also be bypassed.

\subhead{Kernel version}
Since \ourframework{} relies on specific eBPF functionality, devices with kernel version 5.10 or higher, such as Google Pixel 6 and newer, are required. Although \gls{ebpf} can also run on emulators, kernel version remains critical for both devices and emulators, as different versions support varying features that impact effectiveness. Ensuring compatibility between the kernel and the required eBPF features is therefore essential to avoid limitations or inconsistent behavior.

\subhead{Conditional anti-analysis checks}
Identifying anti-analysis tricks was challenging, as some crashes occurred silently without clear \texttt{logcat} evidence, and our tests only covered app launch and initial permissions. This limitation is common in dynamic analysis, where protections may be triggered only through specific user interactions or delayed routines. Since our runtime analysis was restricted to 15 seconds, certain defenses may have gone undetected, making the results in Section~\ref{prevalence-analysis} a conservative lower bound.

\subhead{Java-based anti-analysis checks}
\ourframework{} is limited in fully Java-based APIs manipulation due to parsing ART functions and stack unwinding as \gls{ebpf} programmability is very limited and has to pass verifier. We consider this to be future work.

\subhead{Non-writable memory regions} In general, writing to memory regions that are not flagged as writable is a limitation of \ourframework{}. Nonetheless, by combining runtime offset calculation with the use of \texttt{mprotect} events, it becomes possible to write to memory locations that were initially loaded with \texttt{rw-} permissions and later changed to read-only regions.

\section{Related Work}
There has been a vast collection of published studies on packing/unpacking efforts as well as related analyses and surveys. In this section, we briefly discuss several. Refer to Section~\ref{sec:unpackers} for a survey of unpackers.

%Prior works have explored various strategies to analyze or bypass protections in packed apps. 
To achieve packer evasion, Davinci~\cite{druffel2020davinci} uses a kernel module to hook specific system calls and conceal analysis artifacts, but suffers from maintainability issues and limited flexibility as they only manipulate syscall return values. Rasthofer et al.~\cite{rasthofer2016harvesting} addressed emulator detection with Dex code slicing, but their approach remains ineffective against native-code anti-analysis techniques.
Ruggia et al.~\cite{ruggia2024unmasking} propose DroidDungeon, a sandbox with stack unwinding, but its reliance on fixed anti-evasion lists limits adaptability. PackDiff~\cite{packdiff} applies multi-layer instrumentation for differential analysis but is constrained by Android versioning and its only focus on free packing services. Finally, NCScope~\cite{zhou2022ncscope} combines hardware tracing (ETM\footnote{Embedded Trace Macrocell, an optional hardware component in ARM processors that provides real-time instruction and data tracing.}) with \gls{ebpf} for semantic analysis, though its dependency on specialized hardware restricts usability. For identifying anti-analysis tricks, Sue et al.~\cite{suo2025arap} manually created fingerprints and applied them to detect similar techniques in other apps.

\section{Conclusion}
%Its an arms-race that the analysis side were losing the fight. 
By confirming the high impact of Android packers with a prevalence analysis (particularly on Chinese apps), and the ineffectiveness of current unpackers with an experimental survey, this paper proposed a paradigm shift to create an evasion framework that exists alongside packers, to still allow tools like Frida to analyze what remains available at runtime. We designed and implemented \ourframework{} using eBPF that takes evasion rules (DERs) as input to bypass anti-analysis techniques of the packers. We evaluated \ourframework{} by measuring the number of additional apps (for which we have the known DERs)
and more importantly, the extent to which past dynamic analysis works can benefit from \ourframework{}, e.g., we were able to see significantly more device fingerprints for a fingerprinting detection paper and access operations for an SDK covert channel detection paper, compared to when running without \ourframework{}. Our work sheds light on future research to achieve a balance between privacy, security, and intellectual property.

% \clearpage
\balance
\bibliographystyle{plain}
\bibliography{references}

\begin{thebibliography}{10}

\bibitem{appdome}
Appdome – ai-native protection for the mobile business.
\newblock \url{https://www.appdome.com/}.
\newblock Accessed: 2025-08-12.

\bibitem{doverunner}
Doverunner – complete mobile application and content security.
\newblock \url{https://doverunner.com/}.
\newblock Accessed: 2025-08-12.

\bibitem{firebase}
Firebase — google’s mobile and web app development platform.
\newblock \url{https://firebase.google.com/}.
\newblock Accessed: 2025-08-13.

\bibitem{frida_dexdump_hluwa}
hluwa/frida-dexdump: A frida tool to dump dex in memory for malware analysis.
\newblock \url{https://github.com/hluwa/frida-dexdump}.
\newblock Accessed: 2025-08-15.

\bibitem{play_integrity_api}
Play integrity api – google play security and integrity services.
\newblock \url{https://developer.android.com/google/play/integrity}.
\newblock Accessed: 2025-08-13.

\bibitem{objection_sensepost}
sensepost/objection: Runtime mobile exploration toolkit powered by frida.
\newblock \url{https://github.com/sensepost/objection}.
\newblock Accessed: 2025-08-15.

\bibitem{lemon_eurecom_s3}
eurecom-s3/lemon: Lemon – an ebpf memory dump tool for x64 and arm64 linux and android.
\newblock \url{https://github.com/eurecom-s3/lemon}, 2025.
\newblock Accessed: 2025-08-19.

\bibitem{360Dev}
{360 Developer}.
\newblock 360 developer platform, 2025.
\newblock Accessed: 2025-02-09.

\bibitem{so_app}
{360 Security Technology Inc.}
\newblock So app - mobile security and utility application.
\newblock \url{https://app.so.com/}, 2025.
\newblock Accessed: 2024-05-23.

\bibitem{aldoseri2022insecure}
Abdulla Aldoseri and David Oswald.
\newblock insecure:: Vulnerability analysis of uri scheme handling in android mobile browsers.
\newblock In {\em Workshop on Measurements, Attacks, and Defenses for the Web (MADWeb) 2022}. The Internet Society, 2022.

\bibitem{alecci2024appCategorization}
Marco Alecci, Jordan Samhi, Tegawende~F. Bissyande, and Jacques Klein.
\newblock Revisiting android app categorization.
\newblock In {\em Proceedings of the IEEE/ACM 46th International Conference on Software Engineering}, ICSE '24, New York, NY, USA, 2024. Association for Computing Machinery.

\bibitem{anglano2023enabling}
Cosimo Anglano, Massimo Canonico, Andrea Cepollina, Davide Freggiaro, Alderico Gallo, and Marco Guazzone.
\newblock Enabling the forensic study of application-level encrypted data in android via a frida-based decryption framework.
\newblock In {\em Proceedings of the 18th International Conference on Availability, Reliability and Security}, pages 1--10, 2023.

\bibitem{apkid}
APKiD.
\newblock \url{https://github.com/rednaga/APKiD}, 2025.

\bibitem{apkpure_website}
{APKPure}.
\newblock Apkpure - free and safe android apk downloads.
\newblock \url{https://apkpure.com/}, 2025.
\newblock Accessed: 2024-04-20.

\bibitem{arp2014drebin}
Daniel Arp, Michael Spreitzenbarth, Malte Hubner, Hugo Gascon, Konrad Rieck, and CERT Siemens.
\newblock Drebin: Effective and explainable detection of android malware in your pocket.
\newblock In {\em Ndss}, volume~14, pages 23--26, 2014.

\bibitem{aya_rs_website}
{Aya Rust}.
\newblock Aya - a rust library for ebpf and bpf type format (btf) in the linux kernel.
\newblock \url{https://aya-rs.dev/}, 2025.
\newblock Accessed: 2025-01-08.

\bibitem{BaiduApp}
{Baidu}.
\newblock Baidu app, 2025.
\newblock Accessed: 2025-02-09.

\bibitem{Bangcle}
{Bangcle}.
\newblock Bangcle security solutions, 2025.
\newblock Accessed: 2025-02-09.

\bibitem{barros2015static}
Paulo Barros, Ren{\'e} Just, Suzanne Millstein, Paul Vines, Werner Dietl, Marcelo d'Amorim, and Michael~D Ernst.
\newblock Static analysis of implicit control flow: Resolving java reflection and android intents (t).
\newblock In {\em 2015 30th IEEE/ACM International Conference on Automated Software Engineering (ASE)}, pages 669--679. IEEE, 2015.

\bibitem{bayer2009view}
Ulrich Bayer, Imam Habibi, Davide Balzarotti, Engin Kirda, and Christopher Kruegel.
\newblock A view on current malware behaviors.
\newblock In {\em LEET}, 2009.

\bibitem{berlato2020jisa}
Stefano Berlato and Mariano Ceccato.
\newblock A large-scale study on the adoption of anti-debugging and anti-tampering protections in android apps.
\newblock {\em Journal of Information Security and Applications}, 52:102463, 2020.

\bibitem{cai2020redex}
Jiajin Cai, Tongxin Li, Can Huang, and Xinhui Han.
\newblock Redex: Unpacking android packed apps by executing every method.
\newblock In {\em 2020 IEEE 19th International Conference on Trust, Security and Privacy in Computing and Communications (TrustCom)}, pages 337--344. IEEE, 2020.

\bibitem{medusa_ch0pin}
Ch0pin.
\newblock medusa: A binary instrumentation framework based on frida.
\newblock GitHub repository, \url{https://github.com/Ch0pin/medusa}, 2025.
\newblock Accessed: 2025-08-15.

\bibitem{chame1eon_jnitrace}
{chame1eon}.
\newblock jnitrace: A frida-based tool that traces usage of the jni api in android apps.
\newblock \url{https://github.com/chame1eon/jnitrace}, 2025.
\newblock Accessed: 2025-05-24.

\bibitem{cheng2021obfuscation}
Binlin Cheng, Jiang Ming, Erika~A Leal, Haotian Zhang, Jianming Fu, Guojun Peng, and Jean-Yves Marion.
\newblock $\{$Obfuscation-Resilient$\}$ executable payload extraction from packed malware.
\newblock In {\em 30th USENIX Security Symposium (USENIX Security 21)}, pages 3451--3468, 2021.

\bibitem{anti_debugging_android}
Haehyun Cho, Jongsu Lim, Hyunki Kim, and Jeong~Hyun Yi.
\newblock Anti-debugging scheme for protecting mobile apps on android platform.
\newblock {\em J. Supercomput.}, 72(1):232–246, January 2016.

\bibitem{cocos_website}
{Cocos}.
\newblock Cocos - cross-platform game development engine.
\newblock \url{https://www.cocos.com/en}, 2025.
\newblock Accessed: 2025-01-08.

\bibitem{blackdex_github}
{CodingGay}.
\newblock Blackdex - runtime decryption of android applications.
\newblock \url{https://github.com/CodingGay/BlackDex/}, 2025.
\newblock Accessed: 2025-01-08.

\bibitem{Cui2022TraceDroid}
Huajun Cui, Guozhu Meng, Yan Zhang, Weiping Wang, Dali Zhu, Ting Su, Xiaodong Zhang, and Yuejun Li.
\newblock Tracedroid: A robust network traffic analysis framework for privacy leakage in android apps.
\newblock In {\em Science of Cyber Security: 4th International Conference, SciSec 2022, Matsue, Japan, August 10–12, 2022, Revised Selected Papers}, page 541–556, Berlin, Heidelberg, 2022. Springer-Verlag.

\bibitem{diamantaris2019reaper}
Michalis Diamantaris, Elias~P Papadopoulos, Evangelos~P Markatos, Sotiris Ioannidis, and Jason Polakis.
\newblock Reaper: real-time app analysis for augmenting the android permission system.
\newblock In {\em Proceedings of the Ninth ACM Conference on Data and Application Security and Privacy}, pages 37--48, 2019.

\bibitem{packdiff}
Zikan Dong, Hongxuan Liu, Liu Wang, Xiapu Luo, Yao Guo, Guoai Xu, Xusheng Xiao, and Haoyu Wang.
\newblock What did you pack in my app? a systematic analysis of commercial android packers.
\newblock In {\em Proceedings of the 30th ACM Joint European Software Engineering Conference and Symposium on the Foundations of Software Engineering}, ESEC/FSE 2022, page 1430–1440, New York, NY, USA, 2022. Association for Computing Machinery.

\bibitem{dong2024exploring}
Zikan Dong, Tianming Liu, Jiapeng Deng, Li~Li, Minghui Yang, Meng Wang, Guosheng Xu, and Guoai Xu.
\newblock Exploring covert third-party identifiers through external storage in the android new era.
\newblock In {\em 33rd USENIX Security Symposium (USENIX Security 24)}, pages 4535--4552, 2024.

\bibitem{druffel2020davinci}
Alexander Druffel and Kris Heid.
\newblock Davinci: Android app analysis beyond frida via dynamic system call instrumentation.
\newblock In {\em Applied Cryptography and Network Security Workshops: ACNS 2020 Satellite Workshops, AIBlock, AIHWS, AIoTS, Cloud S\&P, SCI, SecMT, and SiMLA, Rome, Italy, October 19--22, 2020, Proceedings 18}, pages 473--489. Springer, 2020.

\bibitem{Duan2018ThingsYM}
Yue Duan, Mu~Zhang, Abhishek~Vasisht Bhaskar, Heng Yin, Xiaorui Pan, Tongxin Li, Xueqiang Wang, and XiaoFeng Wang.
\newblock Things you may not know about android (un)packers: A systematic study based on whole-system emulation.
\newblock In {\em Network and Distributed System Security Symposium}, 2018.

\bibitem{ebpf_website}
{eBPF.io}.
\newblock ebpf - extended berkeley packet filter.
\newblock \url{https://ebpf.io}, 2025.
\newblock Accessed: 2025-01-08.

\bibitem{unreal_mobile}
{Epic Games}.
\newblock Unreal engine mobile games - develop, optimize, and scale mobile games.
\newblock \url{https://www.unrealengine.com/en-US/uses/mobile-games}, 2025.
\newblock Accessed: 2025-01-08.

\bibitem{filecentipede}
{filecxx}.
\newblock File centipede: All-in-one internet file upload/download manager.
\newblock \url{https://filecxx.com/en_US/index.html}, 2025.
\newblock Accessed: 2025-05-08.

\bibitem{frida_website}
{Frida}.
\newblock Frida - dynamic instrumentation toolkit for developers, reverse engineers, and security researchers.
\newblock \url{https://frida.re/}, 2025.
\newblock Accessed: 2025-01-08.

\bibitem{geffner_ssl_logger}
Jason Geffner.
\newblock ssl\_logger: Decrypts and logs a process's ssl traffic.
\newblock \url{https://github.com/google/ssl_logger}, 2015.
\newblock Archived by Google on December 29, 2022. Accessed: 2025-05-24.

\bibitem{flutter_website}
{Google Developers}.
\newblock Flutter - build apps for any screen.
\newblock \url{https://flutter.dev/}, 2025.
\newblock Accessed: 2025-01-08.

\bibitem{graux2019obfuscated}
Pierre Graux, Jean-Fran{\c{c}}ois Lalande, and Val{\'e}rie Viet~Triem Tong.
\newblock Obfuscated android application development.
\newblock In {\em Proceedings of the Third Central European Cybersecurity Conference}, pages 1--6, 2019.

\bibitem{guo2025code}
Jiawei Guo, Yu~Nong, Zhiqiang Lin, and Haipeng Cai.
\newblock Code speaks louder: Exploring security and privacy relevant regional variations in mobile applications.
\newblock In {\em IEEE Symposium on Security and Privacy (S\&P)}, pages 3952--3970, 2025.

\bibitem{fart_github}
{hanbinglengyue}.
\newblock Fart - a framework for android reverse engineering and taint analysis.
\newblock \url{https://github.com/hanbinglengyue/FART/tree/master}, 2025.
\newblock Accessed: 2025-01-08.

\bibitem{haupert2018honey}
Vincent Haupert, Dominik Maier, Nicolas Schneider, Julian Kirsch, and Tilo M{\"u}ller.
\newblock Honey, i shrunk your app security: The state of android app hardening.
\newblock In {\em International Conference on Detection of Intrusions and Malware, and Vulnerability Assessment}, pages 69--91. Springer, 2018.

\bibitem{heid2024haven}
Kris Heid and Jens Heider.
\newblock Haven't we met before?-detecting device fingerprinting activity on android apps.
\newblock In {\em Proceedings of the 2024 European Interdisciplinary Cybersecurity Conference}, pages 11--18, 2024.

\bibitem{ibrahim2021safetynot}
Muhammad Ibrahim, Abdullah Imran, and Antonio Bianchi.
\newblock Safetynot: on the usage of the safetynet attestation api in android.
\newblock In {\em Proceedings of the 19th Annual International Conference on Mobile Systems, Applications, and Services}, pages 150--162, 2021.

\bibitem{iJiami}
{iJiami}.
\newblock ijiami security solutions, 2025.
\newblock Accessed: 2025-02-09.

\bibitem{magisk_github}
{John Wu}.
\newblock Magisk - a modern rooting solution for android.
\newblock \url{https://github.com/topjohnwu/Magisk}, 2025.
\newblock Accessed: 2025-01-08.

\bibitem{koch2023ok}
Simon Koch, Benjamin Altpeter, and Martin Johns.
\newblock The $\{$OK$\}$ is not enough: A large scale study of consent dialogs in smartphone applications.
\newblock In {\em 32nd USENIX Security Symposium (USENIX Security 23)}, pages 5467--5484, 2023.

\bibitem{kondracki2022droid}
Brian Kondracki, Babak~Amin Azad, Najmeh Miramirkhani, and Nick Nikiforakis.
\newblock The droid is in the details: Environment-aware evasion of android sandboxes.
\newblock In {\em Proceedings of the 29th Network and Distributed System Security Symposium (NDSS)}, 2022.

\bibitem{li2025bpfdex}
Mingyang Li, Weina Niu, Jiacheng Gong, Song Li, Mingxue Zhang, and Xiaosong Zhang.
\newblock Bpfdex: Enabling robust android apps unpacking via android kernel.
\newblock {\em IEEE Transactions on Information Forensics and Security}, 2025.

\bibitem{liu2024tickets}
Yijing Liu, Yiming Zhang, Baojun Liu, Haixin Duan, Qiang Li, Mingxuan Liu, Ruixuan Li, and Jia Yao.
\newblock Tickets or privacy? understand the ecosystem of chinese ticket grabbing apps.
\newblock In {\em 33rd USENIX Security Symposium (USENIX Security 24)}, pages 5107--5124, 2024.

\bibitem{liapp_website}
{Lockin Company}.
\newblock Liapp - mobile application security and anti-tampering solution.
\newblock \url{https://liapp.lockincomp.com/}, 2025.
\newblock Accessed: 2025-01-08.

\bibitem{lsposed_github}
{LSPosed Team}.
\newblock Lsposed - a riru/enhanced xposed module for android.
\newblock \url{https://github.com/LSPosed/LSPosed}, 2025.
\newblock Accessed: 2025-01-08.

\bibitem{Manxi}
{Manxi Inc.}
\newblock Manxi inc., 2025.
\newblock Accessed: 2025-02-09.

\bibitem{notrepackage}
Alessio Merlo, Antonio Ruggia, Luigi Sciolla, and Luca Verderame.
\newblock You shall not repackage! demystifying anti-repackaging on android.
\newblock {\em Computers \& Security}, 103:102181, 2021.

\bibitem{selenium_python_bindings}
Baiju Muthukadan.
\newblock Selenium with python.
\newblock \url{https://selenium-python.readthedocs.io/}, 2025.
\newblock Selenium Python Bindings Documentation.

\bibitem{npmanager}
normalplayer.
\newblock Np manager.
\newblock \url{http://normalplayer.top/}, 2024.

\bibitem{owasp_masvs}
{OWASP Foundation}.
\newblock Mobile application security verification standard (masvs).
\newblock \url{https://mas.owasp.org/MASVS/}, 2025.
\newblock Accessed: 2025-08-03.

\bibitem{certvalidation2024Sajjad}
Sajjad Pourali, Xiufen Yu, Lianying Zhao, Mohammad Mannan, and Amr Youssef.
\newblock Racing for tls certificate validation: a hijacker's guide to the android tls galaxy.
\newblock In {\em Proceedings of the 33rd USENIX Conference on Security Symposium}, SEC '24, USA, 2024. USENIX Association.

\bibitem{rasthofer2016harvesting}
Siegfried Rasthofer, Steven Arzt, Marc Miltenberger, and Eric Bodden.
\newblock Harvesting runtime values in android applications that feature anti-analysis techniques.
\newblock In {\em NDSS}, 2016.

\bibitem{rice2023learning}
Liz Rice.
\newblock {\em Learning eBPF}.
\newblock " O'Reilly Media, Inc.", 2023.

\bibitem{ruggia2024unmasking}
Antonio Ruggia, Dario Nisi, Savino Dambra, Alessio Merlo, Davide Balzarotti, and Simone Aonzo.
\newblock Unmasking the veiled: A comprehensive analysis of android evasive malware.
\newblock In {\em Proceedings of the 19th ACM Asia Conference on Computer and Communications Security}, pages 383--398, 2024.

\bibitem{sahin2018instruction}
Onur Sahin, Ayse~K. Coskun, and Manuel Egele.
\newblock Proteus: Detecting android emulators from instruction-level profiles.
\newblock In Michael Bailey, Thorsten Holz, Manolis Stamatogiannakis, and Sotiris Ioannidis, editors, {\em Research in Attacks, Intrusions, and Defenses}, pages 3--24, Cham, 2018. Springer International Publishing.

\bibitem{strazzere2014android}
T.~Strazzere.
\newblock Android hacker protection level 0, 2014.
\newblock DEF CON 22 Presentation.

\bibitem{dexx2018sun}
Caijun Sun, Hua Zhang, Su-Juan Qin, Nengqiang He, Jiawei Qin, and Hongwei Pan.
\newblock Dexx: A double layer unpacking framework for android.
\newblock {\em IEEE Access}, PP:1--1, 10 2018.

\bibitem{suo2025arap}
Dewen Suo, Lei Xue, Le~Yu, Runze Tan, Weihao Huang, and Guozi Sun.
\newblock Arap: Demystifying anti runtime analysis code in android apps.
\newblock {\em IEEE Transactions on Software Engineering}, 2025.

\bibitem{sutter2024dynamic}
Thomas Sutter, Timo Kehrer, Marc Rennhard, Bernhard Tellenbach, and Jacques Klein.
\newblock Dynamic security analysis on android: A systematic literature review.
\newblock {\em IEEE Access}, 12:57261--57287, 2024.

\bibitem{TencentCloud}
{Tencent Cloud}.
\newblock Tencent cloud, 2025.
\newblock Accessed: 2025-02-09.

\bibitem{turner2025android}
{Turner, Ash}.
\newblock Android vs.\ apple market share: Leading mobile operating systems (os), January 2025.
\newblock BankMyCell; updated January4,2025; accessed August 3,2025.

\bibitem{ugarte2015sok}
Xabier Ugarte-Pedrero, Davide Balzarotti, Igor Santos, and Pablo~G Bringas.
\newblock Sok: Deep packer inspection: A longitudinal study of the complexity of run-time packers.
\newblock In {\em 2015 IEEE Symposium on Security and Privacy}, pages 659--673. IEEE, 2015.

\bibitem{unity_mobile}
{Unity Technologies}.
\newblock Unity mobile solutions - build, operate, and grow mobile games and apps.
\newblock \url{https://unity.com/solutions/mobile}, 2025.
\newblock Accessed: 2025-01-08.

\bibitem{beyondgoogle2018wang}
Haoyu Wang, Zhe Liu, Jingyue Liang, Narseo Vallina-Rodriguez, Yao Guo, Li~Li, Juan Tapiador, Jingcun Cao, and Guoai Xu.
\newblock Beyond google play: A large-scale comparative study of chinese android app markets.
\newblock In {\em Proceedings of the Internet Measurement Conference 2018}, IMC '18, page 293–307, New York, NY, USA, 2018. Association for Computing Machinery.

\bibitem{blutter_github}
Worawit Wang.
\newblock Blutter - a linux kernel exploit framework.
\newblock \url{https://github.com/worawit/blutter}, 2025.
\newblock Accessed: 2025-01-08.

\bibitem{wong2018tackling}
Michelle~Y Wong and David Lie.
\newblock Tackling runtime-based obfuscation in android with $\{$TIRO$\}$.
\newblock In {\em 27th USENIX security symposium (USENIX security 18)}, pages 1247--1262, 2018.

\bibitem{packergrind2017leixue}
Lei Xue, Xiapu Luo, Le~Yu, Shuai Wang, and Dinghao Wu.
\newblock Adaptive unpacking of android apps.
\newblock In {\em 2017 IEEE/ACM 39th International Conference on Software Engineering (ICSE)}, pages 358--369, 2017.

\bibitem{xue2021parema}
Lei Xue, Yuxiao Yan, Luyi Yan, Muhui Jiang, Xiapu Luo, Dinghao Wu, and Yajin Zhou.
\newblock Parema: an unpacking framework for demystifying vm-based android packers.
\newblock In {\em Proceedings of the 30th ACM SIGSOFT International Symposium on Software Testing and Analysis}, pages 152--164, 2021.

\bibitem{happer2021unpacker}
Lei Xue, Hao Zhou, Xiapu Luo, Yajin Zhou, Yang Shi, Guofei Gu, Fengwei Zhang, and Man~Ho Au.
\newblock Happer: Unpacking android apps via a hardware-assisted approach.
\newblock In {\em 2021 IEEE Symposium on Security and Privacy (SP)}, pages 1641--1658, 2021.

\bibitem{Yang2024BeyondHorizon}
Shishuai Yang, Guangdong Bai, Ruoyan Lin, Jialong Guo, and Wenrui Diao.
\newblock Beyond the horizon: Exploring cross-market security discrepancies in parallel android apps.
\newblock In {\em 2024 IEEE 35th International Symposium on Software Reliability Engineering (ISSRE)}, pages 558--569, 2024.

\bibitem{appspear2015yang}
Wenbo Yang, Yuanyuan Zhang, Juanru Li, Junliang Shu, Bodong Li, Wenjun Hu, and Dawu Gu.
\newblock Appspear: Bytecode decrypting and dex reassembling for packed android malware.
\newblock In {\em Proceedings of the 18th International Symposium on Research in Attacks, Intrusions, and Defenses - Volume 9404}, RAID 2015, page 359–381, Berlin, Heidelberg, 2015. Springer-Verlag.

\bibitem{unpacker_github}
{Youlor}.
\newblock Unpacker - a tool for extracting and analyzing packed android applications.
\newblock \url{https://github.com/youlor/unpacker}, 2025.
\newblock Accessed: 2025-01-08.

\bibitem{yu2014android}
Rowland Yu.
\newblock Android packers: facing the challenges, building solutions.
\newblock In {\em Proceedings of the 24th Virus Bulletin International Conference}, 2014.

\bibitem{zerbini2024r}
Simone Zerbini, Samuele Doria, Primal Wijesekera, Serge Egelman, and Eleonora Losiouk.
\newblock R+ r: Matrioska: A user-centric defense against virtualization-based repackaging malware on android.
\newblock In {\em 2024 Annual Computer Security Applications Conference (ACSAC)}, pages 843--856. IEEE, 2024.

\bibitem{zhang2015dexhunter}
Y.~Zhang, X.~Luo, and H.~Yin.
\newblock Dexhunter: Toward extracting hidden code from packed android applications.
\newblock In {\em Computer Security -- ESORICS 2015: 20th European Symposium on Research in Computer Security, Vienna, Austria, September 21-25, 2015, Proceedings, Part II}, pages 293--311. Springer, 2015.

\bibitem{zheng2025gupacker}
Tao Zheng, Qiyu Hou, Xingshu Chen, Hao Ren, Meng Li, Hongwei Li, and Changxiang Shen.
\newblock Gupacker: Generalized unpacking framework for android malware.
\newblock {\em IEEE Transactions on Information Forensics and Security}, 2025.

\bibitem{zhou2022ncscope}
Hao Zhou, Shuohan Wu, Xiapu Luo, Ting Wang, Yajin Zhou, Chao Zhang, and Haipeng Cai.
\newblock Ncscope: hardware-assisted analyzer for native code in android apps.
\newblock In {\em Proceedings of the 31st ACM SIGSOFT International Symposium on Software Testing and Analysis}, pages 629--641, 2022.

\end{thebibliography}
\clearpage
%\section*{Appendix}

% \usepackage{listings}
\vspace{4em}
\section{Syscall-region Mapping}\label{appI}
Assisted analysis – showing candidate anti-analysis syscalls that are originated
from the same code region (e.g. \tt{/proc/self/wchan}).
\lstset{
  basicstyle=\ttfamily\scriptsize,
  breaklines=true,
  columns=fullflexible
}

\begin{lstlisting}
[region 94] 0x7a41c45000 - 0x7a41d235c0 (r-x)
openat(*pathname=0x7a41d2d030(/proc/self/status))
openat(*pathname=0x7a41d2d090(/proc/self/wchan))
openat(*pathname=0x7a41d2a588(/proc/self/maps))
openat(*pathname=0x7fdc6f19e0(/apex/com.android.art
/lib64/libart.so))
openat(*pathname=0x7fdc6f1a10(/apex/com.android.art
/lib64/libart.so))
openat(*pathname=0x7fdc6f1a10(/system/lib64/liblog.so))
openat(*pathname=0x7a3d701090(/apex/com.android.art
/lib64/libart.so))
openat(*pathname=0x7a3d505080(/apex/com.android.run-
time/bin/linker64))
openat(*pathname=0x7a41d2d0e4(/proc/self/maps))
openat(*pathname=0x7fdc6f17b0(/apex/com.android.art
/lib64/libart.so))
openat(*pathname=0x7fdc6f17b0(/apex/com.android.art
/lib64/libart.so))
openat(*pathname=0x7fdc6f1ba0(/apex/com.android.art
/lib64/libart.so))
openat(*pathname=0x7a41d2d030(/proc/self/status))
[region 95] 0x7d79b34000 - 0x7d79b35000 (---)
\end{lstlisting}

\section{Distribution of Packers Across Tiers}
\label{packer-tiers}
\begin{CJK*}{UTF8}{gbsn}

\begin{table}[H]
  \centering
  \scriptsize
  \setlength{\tabcolsep}{3pt}
  \renewcommand{\arraystretch}{1.08}
  \begin{tabularx}{\columnwidth}{@{}l l r@{}}
    \toprule
    \textbf{Packer} & \textbf{Variant} & \textbf{\#apps (\%)} \\
    \midrule
    \multirow{3}{*}{360付费版 (360 Security)}
      & 360加固 (Basic) & 2638 (33.33\%) \\
      & 360加固企业版 (Enterprise) & 44 (0.55\%) \\
      & 360付费版 (Paid) & 18 (0.22\%) \\
    \midrule
    \multirow{2}{*}{梆梆加固 (Bangcle)}
      & 梆梆加固 (Basic) & 95 (1.2\%) \\
      & 梆梆加固企业版 (Enterprise) & 352 (4.4\%) \\
    \midrule
    \multirow{2}{*}{爱加密 (Ijiami)}
      & 爱加密 (Basic) & 82 (1.03\%) \\
      & 爱加密企业版 (Enterprise) & 312 (3.94\%) \\
    \midrule
    \multirow{3}{*}{腾讯御安全 (Tencent)}
      & 腾讯御安全 (Basic) & 218 (2.75\%) \\
      & 腾讯御安全(旧) (Legacy) & 246 (3.10\%) \\
      & 腾讯御安全企业版 (Enterprise) & 11 (3.10\%) \\
    \bottomrule
  \end{tabularx}
  \caption{NP-Manager packer tier identification (Chinese apps).}
  \label{tab:packer_tier_results}
\end{table}

\end{CJK*}

\section{Packers Loading Native Library} \label{app:native-loader}
Packers load their libraries at app's launch using Java function calls to \texttt{System.load()} or \texttt{System.loadLibrary()}, typically within the \texttt{onCreate()} method. Runtime library loading is handled by the Android loader, located at \texttt{/apex/com.android.runtime/bin/linker64}. Packers can define custom library constructors, which are invoked by symbols like \texttt{do\_dlopen()} or \texttt{call\_constructor()} in \texttt{linker64}. Additionally, while \gls{jni} functions typically initialize in \texttt{JNI\_OnLoad}, another function, \texttt{init\_proc}, executes beforehand. To hook into these functions, security researchers must first intercept the Android loader and exclude other library initializations.

\section{Packer Code Release Stages} \label{sec:stages}

\begin{figure}[H]
    \centering
    \includegraphics[width=\columnwidth]{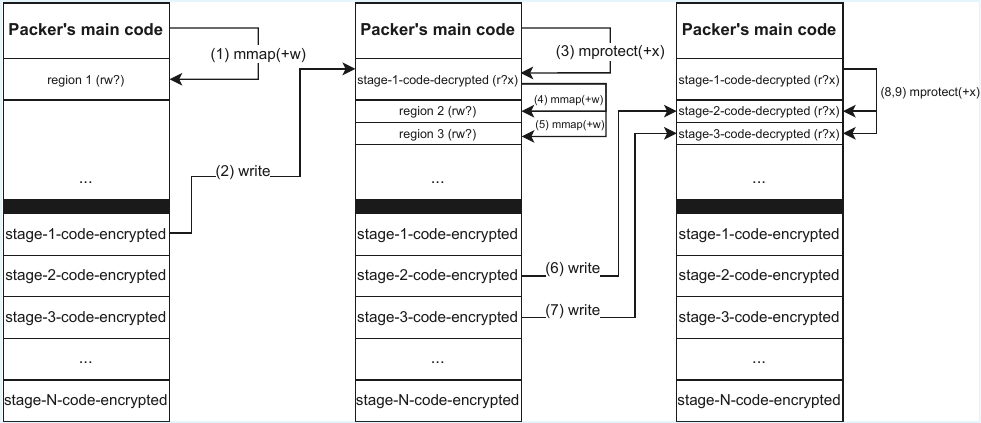}
    \caption[Packer multi-stage code release overview.]
    {Packer multi-stage code release overview. Shows how packers decrypt their 
    code for next stages at runtime. (?): Whether the permission is enabled or 
    not (rwx), (\#): Dynamic code loading step numbers.}
    \label{fig:packer-stages}
\end{figure}

% \section{eBPF multiple instances}
% \section{diff analysis}

\end{document}